\documentclass[letterpaper,12pt]{article}
\usepackage[top=1in, bottom=1in, left=1in, right=1in]{geometry}
\usepackage{setspace}
\usepackage{mathrsfs}
\usepackage{dsfont}
\usepackage{amsmath,amsthm,amsfonts,amssymb,amscd}
\usepackage{eqnarray}
\usepackage{mdframed}
\mdfdefinestyle{codebox}{
    backgroundcolor=gray!10,
    linecolor=gray!60,
    linewidth=0.5pt,
    roundcorner=3pt,
    innerleftmargin=8pt,
    innerrightmargin=8pt,
    innertopmargin=6pt,
    innerbottommargin=6pt
}

\usepackage{upquote}    

\usepackage[T1]{fontenc}
\usepackage{lmodern}

\usepackage{multirow}

\usepackage{multicol}
\usepackage{soul}
\usepackage{xcolor}
\sethlcolor{yellow} 

\newenvironment{hypothesis}[1]{%
  \par\noindent\textbf{Hypothesis #1:} \itshape%
}{%
  \par%
}

\usepackage[pdftex]{hyperref}
\hypersetup{colorlinks, citecolor=black, filecolor=blue, linkcolor=blue, urlcolor=blue}

\usepackage{booktabs} 
\usepackage{longtable} 
\usepackage{pdflscape} 
\usepackage{array} 
\usepackage{ragged2e}
\usepackage{url}

\usepackage[table]{xcolor} 
\definecolor{vcDarkGreen}{RGB}{0,88,52} 
\definecolor{vcDarkBrownYellow}{RGB}{145,105,25}
\definecolor{vcYellowBrown}{RGB}{181,145,64} \definecolor{vcLightRed}{RGB}{185,96,96} \definecolor{vcDarkRed}{RGB}{128,30,30}

\usepackage{longtable}
\usepackage{csquotes}
\usepackage{caption}
\usepackage{subcaption}

\usepackage{array} 
\usepackage{ragged2e}
\usepackage{url}

\usepackage{amsmath}
\usepackage{booktabs} 
\newcommand{\tabnotes}[2]{\bottomrule \multicolumn{#1}{@{}p{0.95\linewidth}@{}}{\footnotesize #2 }\end{tabular}\end{table}}
\usepackage{caption}
\usepackage{float} 
\usepackage{graphicx} 
\usepackage[toc,page]{appendix} 
\usepackage{listings}
\usepackage{tabularx}
\usepackage{rotating} 

\usepackage{standalone}
\usepackage{pdflscape}
\usepackage{blkarray}

\usepackage{apacite}
\usepackage[round]{natbib}

\newcommand{\sym}[1]{#1}

\usepackage{titlesec}

\titlespacing*{\section}{0pt}{1.5ex plus 1ex minus .2ex}{0.8ex plus .2ex}
\titlespacing*{\subsection}{0pt}{1.2ex plus 1ex minus .2ex}{0.8ex plus .2ex}

\titleformat*{\section}{\large\bfseries}
\titleformat*{\subsection}{\normalsize\bfseries}

\title{%
  AI Coding Tools and Digital Entrepreneurship: The Role of Software Expertise\thanks{We acknowledge Bright Data’s Bright Initiative for partially funding the data collection.}
}
\author{
  Ruiqing Cao\thanks{Stockholm School of Economics, \href{mailto:sam.cao@hhs.se}{sam.cao@hhs.se}. Corresponding author.} \and
  Abhishek Bhatia\thanks{Copenhagen Business School, \href{mailto:abbh.si@cbs.dk}{abbh.si@cbs.dk}.}
}
\date{\today}

\begin{document}

\maketitle
\singlespacing

\begin{center}
\end{center}

\begin{abstract}
While digital technologies expand entrepreneurial access by providing technical resources, it is less known whether they enable ventures to create durable value and whether they can substitute for technical expertise accumulated through software work experience. This paper studies how digital ventures respond to the wide diffusion of AI coding tools, and how these responses are shaped by founders’ software expertise. Measuring product-category exposure to AI coding using pre-LLM product descriptions and linking venture launch, traffic, and financing data, we show that exposure to AI coding increases first-time venture launches after 2022Q4, while entrants are less likely to survive but raise more financing conditional on one-year survival. Crucially, founders with software work experience drive a larger fraction of new launches, ameliorate the decline in survival when product development is partially rather than fully automated, and explain all of the increase in venture financing.
\vspace{10em}
\end{abstract}

\pagebreak

\doublespacing

\section{Introduction}
Since mid-2022, advances in large language models (LLMs) started to change how digital products are developed and scaled. Widely adopted AI coding tools allow individuals to generate, modify, and debug entire codebases by describing desired product features to an AI system rather than writing code from scratch \citep{mckinsey2025state}. While AI-generated code has rapidly proliferated worldwide \citep{daniotti2026using}, its impact on digital entrepreneurial ventures is not yet fully understood. Prior research shows that digital technologies increase the quantity and variety of products brought to market \citep{benner2023changing} and help e-commerce ventures create and scale their offerings \citep{kerr2014entrepreneurship,stroube2025mapping}. However, less is known about venture outcomes beyond entry and conditions for when widespread diffusion of technical resources supports viable entrepreneurship instead of merely increase market churn \citep{kerr2009democratizing}.

The widespread diffusion of AI coding tools also raises broader questions about how technological change may replace, augment, or transform the role of expertise \citep{choudhury2020machine,krakowski2023artificial,alekseeva2026artificial,jia2024and,sauermann2026extension}. As AI capabilities expand across tasks, it may substitute for human efforts on generic tasks while making deep expertise more valuable \citep{autor2025expertise,li2026economics}. Despite the importance of founders’ software expertise in a venture's initial resource endowment, relatively little has been concluded about whether software expertise may complement or constrain a venture’s ability to harness the value of increasingly capable AI systems. 

To address these gaps, our study examines two research questions. First, how does AI coding tools affect the entry, survival, and financing of new ventures? Second, how does founders’ software expertise shape ventures’ strategic responses to AI coding tools and their subsequent performance? Drawing on the resource-based view \citep{wernerfelt1984resource,barney1991firm}, ventures’ ability to create and capture value from a resource depends on complementary organizational capabilities \citep{teece1986profiting,bresnahan2002information}. AI may raise the productivity on software tasks, and software-experienced founders are more likely to enter markets in which they can effectively exploit their technological strengths \citep{de2007churn}. Thus, we hypothesize that AI coding tools increase entrepreneurial entry in product categories exposed to these tools -- where AI coding can meaningfully automate at least part of the software development tasks -- and that this increase is larger among founders with software expertise.

On the other hand, AI coding tools may hurt venture survival in exposed product categories, because it is not a scarce resource. Its widespread usage increases market entry and competition, which compresses profits and undermines the economic viability of new ventures. However, as large language models are trained on vast quantities of software code, they can most readily substitute for generic tasks \citep{autor2025expertise} that are well represented in the training data. Therefore, AI may automate part of the product development process \citep{li2026economics}, leaving the rest requiring deeper expertise to human experts, rendering venture survival dependent on software expertise.

Finally, AI coding may enable ventures to directly embed AI capabilities in operations and product offerings, leading to new product development processes and business models beyond mere automation of conventional tasks. These ``AI-native’’ ventures may scale faster, disrupt traditional software businesses \citep{schumpeter1942capitalism}, and attract more venture capital financing. As early adopters and experimenters of generative and agentic AI, software-experienced individuals are particularly likely to accumulate first-hand experience with AI’s capabilities and understand how it applies to specific customer problems. They may also be able to identify which parts of traditional value propositions of a software company can be replaced by emerging AI capabilities, and therefore recognize opportunities for disrupting an industry in which they have prior experience. Therefore, software-experienced founders may be particularly well positioned to design and implement technically robust AI-native ventures.

We test these predictions by combining data on first-time venture launches between 2021Q1 and 2024Q4 from Product Hunt, website traffic from Semrush, external financing from PitchBook, and founders’ prior experience, education, and demographic characteristics from Revelio Labs. To isolate the effects of AI coding tools from those of other AI applications, we use the variation in the inherent characteristics of product offerings that determine their exposure to AI coding, measured from product descriptions. We leverage a frontier open-weights embedding model and agglomerative clustering to partition ventures into distinct product categories, and use LLM-assisted classification to assess the susceptibility of representative products in each category to AI coding and define category-level exposure based on the median product score. The resulting measure captures variation across product categories’ overall exposure \citep{felten2021occupational} to AI coding, rather than across ventures' adoption or usage of specific AI tools which cannot be observed in our data. The validity of the identification strategy relies on an assumption that categories more exposed to AI coding are not also systematically more prone to automation by conversational AI tools (e.g., ChatGPT) and AI image-generation tools (e.g., Midjourney) which entrepreneurs may use in parallel for ideation, marketing, and other activities unrelated to software development.

Our analyses yield three main findings. First, following the widespread diffusion of AI coding tools beginning in 2022Q4, more exposed product categories experienced substantial increases in first-time venture launches. This increase was larger among founders with software experience. Second, entrants in exposed product categories became significantly less likely to survive one year after launch. However, this decline was smaller when founders had software experience and AI coding tools could partially but not fully automate relevant product-development tasks. Third, among ventures that survived for at least one year, those in exposed product categories raised significantly more venture capital financing, particularly when their founders had software experience. Together, these findings provide broad support for our theoretical hypotheses.

Our study contributes to two streams of research in strategy and entrepreneurship. First, we extend research on how digital technologies foster the creation of new products and ventures \citep{kerr2014entrepreneurship,benner2023changing,stroube2025mapping} by studying not only entrepreneurial entry but also post-entry survival and financing following the widespread adoption of AI coding practices. Our findings reveal effects beyond a simple substitution of technical resources: AI coding tools increase product entry and intensify market competition, while also creating fertile ground for disruptive innovation that may challenge traditional software business models through the emergence of more scalable AI-native ventures. Second, our results add to a literature on how technological change alters the complementary value of organizational assets, particularly the knowledge and skills of organizational members \citep{jia2024and,otis2026uneven,trobinger2026experience}. Consistent with the argument that AI automates generic tasks before tasks requiring deeper expertise \citep{autor2025expertise}, our findings suggest that founders’ software expertise remains valuable when AI can perform some, but not all, product-development tasks. More importantly, AI may relocate founders’ effort toward designing and implementing novel value propositions with high scaling potential that embed AI capabilities at the core, which may require deep software expertise and effective adaptation to AI.

\section{Theory} \label{sec:theory}

\subsection{AI Coding Tools and Software Expertise} \label{sec:vibecoding}

AI researcher and engineer Andrej Karpathy coined the colloquial term ``vibecoding’’ in 2025 to describe the emerging practice of \textit{AI coding}, where an individual -- whether a professional software engineer or someone without a technical background -- describes desired product features to an AI system through natural-language prompts, and the system generates or modifies the codebase to implement those features. 
Early tools such as GitHub Copilot and Replit’s ``Generate Code’’ feature primarily generated code within local or cloud-based development environments. Subsequent products, including Replit Agent, Vercel’s v0, Bolt, and Lovable, introduced agentic capabilities, enabling AI systems to prototype interfaces and scaffold complete applications with minimal human intervention. Beginning in mid-2025, the launches of Claude Code and Codex further popularized AI coding agents. Agentic AI systems can orchestrate multi-step software-development workflows toward a goal, and automate related tasks beyond code generation such as conducting Internet searches, accessing and editing local files, executing programs, and calling APIs.

The rise of AI coding tools has led to a sharp increase in AI-generated code \citep{daniotti2026using}. In April 2026, Y Combinator founder Paul Graham stated that, for ``at least a year, maybe two,'' AI had written approximately three quarters of the code produced by startups in each Y Combinator batch. 
AI coding tools transform how engineers work but do not diminish the importance of engineering fundamentals. With the recent wave of agentic AI capabilities, engineers appear to be among the earliest adopters of agentic AI in frontier labs \citep{johnston2026shift}. 
As reflected in a recent Coinbase blog post, \textit{“Getting AI to generate a solution was easy. Generating a good solution, knowing when to trust it, and knowing when to push back required skills”} \citep{coinbase2026interviewing}.

We describe two situations in which non-generic deep software expertise increases the productivity gains from AI coding. The first one is a recent experiment conducted in mid-2026 at Marmelab, an open-source software development studio that builds web, mobile, and AI applications. The experiment shows that selecting the right open-source library and producing high-quality documentation are essential to establishing a solid foundation for cost-effective agentic AI development \citep{rimet2026opensource}:
\begin{quote}
\itshape
\singlespacing
``Popular open-source libraries have been tested by thousands of developers. They have been used in production for years, their maintainers have fixed many bugs and security vulnerabilities, and they publish updates regularly\ldots I recommend building on solid foundations and existing libraries as soon as a project moves beyond a disposable prototype. This is less costly in terms of tokens and yields more reliable results… The benchmark shows that using a well-designed library can reduce both the number of tokens and the time it takes to build an application… Each time an agent uses a well-equipped framework, it therefore weighs less heavily on the AI bill. Open source lowers the price you pay to have an AI write your code, and this benefit grows as agent usage increases… The more features your agent codes without a library, the faster the context fills up in future sessions. Productivity decreases and costs rise as the project grows. Using a library can help you avoid this problem and keep a constant velocity.’’
\end{quote}

The second situation is a discussion by the founder and CEO of Architect Financial Technologies -- a venture-backed fintech and financial-market-infrastructure startup founded in 2023 -- upon LLMs’ inability to build financial trading systems. His reflections suggest that realistic experience and tacit knowledge unavailable to a general-purpose LLM are essential to developing the architectural understanding necessary for creating effective products despite access to frontier AI agents:
\begin{quote}
\itshape
\singlespacing
``My benchmark for testing coding abilities of frontier LLMs is to ask the models to build infrastructure for a market making system from scratch. Even the most recent models, including Fable and GPT-5.6, do not produce systems with design patterns that are commonplace at HFTs. Even though the models’ outputs function in a narrow sense, the code architectures would not be able to form the foundations of even a modest production trading operation\ldots A production trading system is an institutional artifact, a representation of how a firm thinks about markets, failure, ownership, observability, deployment, and risk. Few of the inputs are ever written down in one place, some aren’t written down at all. The knowledge is distributed across code review norms, incident histories, operational procedures, and the accumulated instincts of experienced engineers over decades\ldots This helps explain why public LLMs can’t create effective trading infrastructure from a short specification.''
\end{quote}

\subsection{Hypotheses Development}

\subsubsection{AI Coding Tools and Entrepreneurial Entry}

The resource-based view attributes persistent performance variation across firms to the differences between their resource bundles as well as the differences in their ability to deploy these resources effectively \citep{wernerfelt1984resource,barney1991firm}. Resource constraints can prevent potential entrepreneurs from starting ventures \citep{evans1989estimated} and limit the growth of new ventures \citep{khaire2010young}. A large literature shows that resource provision lowers barriers to entry and increase venture creation \citep{kerr2009democratizing,castellaneta2020intended,dutt2026overcoming}. 
For example, cloud computing increases early-stage software entrepreneurship by reducing the financial capital needed to develop initial products \citep{kerr2014entrepreneurship}, while low-code tools enable the creation and operational scaling of e-commerce ventures by managing the underlying technical infrastructure \citep{stroube2025mapping}. By the same logic, AI coding tools should increase entrepreneurial entry in product categories where they can meaningfully automate some product development tasks.
\begin{hypothesis}{1a (H1a)}
Following the proliferation of AI coding tools, entrepreneurial entry increases in product categories where these tools can autonomously conduct a non-trivial share of product development tasks.
\end{hypothesis}

Resource provision should have larger effects on entrepreneurial entry among populations where it leads to larger expected gains from adoption for venture creation. Potential entrepreneurs differ in the benefits they expect to derive from AI coding tools. Therefore, the extent to which it affects their decision to launch new ventures depends on complementary capabilities that enable the effective deployment of the technical resources that are difficult for rivals to imitate \citep{teece1986profiting,bresnahan2002information}. Because AI is directly relevant to their work activities, software engineers may adopt AI coding tools earlier and integrate it more readily into their workflows. By automating repetitive tasks, they can redirect effort toward activities that require technical expertise, such as system architecture, operational reliability, and codebase maintenance. Therefore, AI coding tools may have a stronger effect on entry among founders with software expertise, who have stronger incentives to engage with AI because they perceive clear and immediate benefits from doing so.

\begin{hypothesis}{1b (H1b)}
Following the proliferation of AI coding tools, founders with software expertise are disproportionately likely to enter product categories where these tools can autonomously conduct a non-trivial share of product development tasks.
\end{hypothesis}

\subsubsection{AI Coding Tools and Survival of Entrepreneurial Entrants}
Widely accessible AI systems do not by themselves confer a competitive advantage on adopters \citep{csaszar2025unbounding}. On the contrary, broad diffusion of AI coding practices may intensify competition and reduce profits in exposed product categories, due to higher entrepreneurial entry and lower demand from users who can use AI themselves to create the same product internally instead of buying it from an external vendor. Recent research shows that the large task-level productivity gains from AI coding tools translate only partially into product releases and do not increase total usage across products \citep{demirer2026writing}, implying that usage declines for the average new product as market entry increases. Therefore, we expect new entrants in product categories more exposed to AI coding practices be become less viable or likely to survive.
\begin{hypothesis}{2a (H2a)}
Following the proliferation of AI coding tools, entrepreneurial entrants in product categories where these tools can conduct a larger share of product development tasks become less likely to survive than earlier entrants in the same categories.
\end{hypothesis}

While founders' technical knowledge and skills have traditionally been crucial to new ventures' product development, they are increasingly codified and automated by AI coding tools. However, such automation may be partial, leaving more context-dependent activities that require deeper knowledge to human experts. Theories of human capital offer a basis for anticipating how AI may reshape the value of expertise. As AI systems learn from vast amounts of codified knowledge, their capabilities may advance more readily in tasks that are more generic and well-documented, for which training data are more abundant. Hence, automation that eliminates generic tasks can increase rather than diminish the value of deep expertise \citep{autor2025expertise,li2026economics}, which is required to complete the remaining parts of a complex project.

In digital ventures, AI coding tools may be unable to fully automate product development across many product categories. For early-stage ventures, survival depends on achieving product-market fit and delivering a product of sufficient quality to satisfy early adopters and generate positive word of mouth. Because product development involves many interdependent tasks that must all be completed successfully, tasks that cannot be automated may become bottlenecks requiring human expertise. For these ventures, software expertise may be crucial to developing viable products that can withstand scrutiny and retain users. We propose:
\begin{hypothesis}{2b (H2b)}
Following the proliferation of AI coding tools, the decline in new venture survival is smaller when founders have software expertise, but only in product categories where these tools partially but not fully automate product development.
\end{hypothesis}

\subsubsection{AI Coding Tools and Financing of Surviving Ventures}
Recent anecdotal evidence suggests that in 2025, AI firms accounted for more than 60\% of the total value of global venture capital (VC) investments almost doubling from 2022, with much of this capital concentrated in large financing rounds \citep{oecd2026venture}. The concentration of venture financing may reflect a close fit between AI ventures and VC investment preferences for innovative firms \citep{hellmann2000interaction}, especially those with the potential for rapid growth and outsized returns despite considerable early-stage uncertainty \citep{lerner2020venture,gompers2020venture}.

Moreover, a pervasive and disruptive technological shift induced by AI may transform industry structures and accelerate creative destruction \citep{schumpeter1942capitalism}, where new business models emerge and threaten to make incumbents obsolete. At a critical junction of technological progress, VCs place risky bets on a small number of firms that they deem as having potential to become category leaders in a transformed industry, such as Google and Amazon in the 2000s and Facebook and Uber in the 2010s. Recent research documents the rise of ``AI-native’’ firms, where AI can enable highly scalable businesses by transforming labor-intensive services into software-based offerings, allowing firms to serve more customers without proportionally increasing headcount \citep{kim2026ai}. In these ventures, AI capabilities reshape organizational processes and become core to product offerings.

We expect surviving ventures in product categories more exposed to AI coding to have a higher likelihood of becoming ``AI-native’’ firms and secure more venture capital financing. Consider \textit{Hoop} and \textit{Legion Health} in Table \ref{tab:vibecodability_examples}, both of which launched in late 2024, several years after the diffusion of AI coding practices. At launch, Hoop was positioned as a fully automated, AI-driven task manager that identified tasks across communication tools and centralized them in one place. In 2026, it pivoted to an AI agent for e-commerce brands that integrates with existing customer-support systems to reduce customer cancellations. Similarly, Legion Health is a full-stack telepsychiatry company that integrates LLMs and AI agents throughout psychiatric care delivery. As an AI-native company, it develops its own clinical software to support AI-agent workflows across functions such as scheduling and billing, clinical decision support, and patient follow-up. In both cases, AI capabilities do not only assist with product development but are also embedded directly into product functions.

Prior research has indeed found that firms investing in AI achieve greater product innovation and attain higher market valuations \citep{babina2024artificial}. When more product components can be automated by AI coding tools rather than created by manually writing code, AI may transform not only the product development process but also the nature of the product offering. AI coding practices may thus enable the creation of novel offerings that embed AI capabilities in the value proposition. We propose: 
\begin{hypothesis}{3a (H3a)}
Following the proliferation of AI coding tools, surviving entrants in product categories where these tools can perform a larger share of product development raise more venture capital financing than earlier surviving entrants in the same categories.
\end{hypothesis}

The scope for product and process innovation should be greater in product categories more exposed to AI coding because a larger share of development work can be reorganized around generative and agentic AI. From a knowledge-based perspective, firms create value by integrating specialized knowledge across their members and applying it to business activities more effectively than markets \citep{conner1996resource,grant1996toward}. Their ability to benefit from AI coding tools may differ, because they may use the same technological resource in markedly different ways and achieve different outcomes \citep{kogut1992knowledge,mcevily2002persistence}. 

Past research on firms’ adoption of information technology shows that complementary organizational assets, particularly skilled human capital, are critical to realizing value from new technologies  \citep{bresnahan2002information,forman2012internet,tambe2014big,brynjolfsson2016rapid,chen2024does}. Because prior expertise provides a foundation for future learning and experimentation, founders with software experience can more readily acquire and productively apply new technical knowledge \citep{dierickx1989asset}, which strengthens their venture’s absorptive capacity related to AI \citep{cohen1990absorptive}. 

Founders' technical knowledge and experimentation with AI across software development tasks help them identify effective ways to redesign entire development workflows that reflect an accurate understanding of what it can and cannot do. This advantage enables their ventures to create value propositions and new product offerings that embed AI capabilities at the core. These ventures, in turn, attract more external financing as venture capital increasingly flows toward AI-native firms. We propose:
\begin{hypothesis}{3b (H3b)}
Following the proliferation of AI coding tools, the increase in external financing among surviving entrants is greater when founders have software expertise, when these tools can autonomously conduct a non-trivial share of product development tasks. 
\end{hypothesis}

\section{Empirical Setting, Data, and Methods} \label{sec:data}

\subsection{Empirical Setting} \label{subsec:setting}
To test the theoretical hypotheses, we combine venture-level data from Product Hunt with auxiliary data sources including Revelio Labs, Semrush, and PitchBook. With roughly four million monthly visits, Product Hunt brings together hundreds of thousands of active users from around the world, including founders of nascent digital startups who launch new products to gauge customer demand signals and gain market traction, and a large community of early adopters who vote and discuss these new launches. Each day, hundreds of digital ventures launch new product by assembling key information about the product, including the venture’s official website URL and a textual product description, along with additional marketing materials such as screenshots and demo videos. 

\subsection{Data and Sample} \label{subsec:sample}
Our empirical data sample consists of digital ventures uniquely identified by their \textit{registered domains}.\footnote{The root domain, rather than subdomains, uniquely identifies each venture. For example, \url{console.berget.ai} and \url{docs.berget.ai} are different subdomains that belong to the same root domain \url{berget.ai} of Berget AI. Ventures without standalone websites (specifically, mobile apps and Chrome extensions) are kept only for product-category level analyses, which we identify using their full third-party URL paths.} We restrict the sample to ventures for which we can observe reliable career histories for the majority of founding team members, which are genuinely new and have not previously launched on Product Hunt or elsewhere.\footnote{See Appendix Section B.2 for detailed criteria to identify entrepreneurial firms undertaking first launch.} All firms included for empirical analyses consist of 41,487 digital ventures first launched between 2021Q1 and 2024Q4, of which 33,155 has standalone websites and are used in venture-level analyses. The final data sample combines several data sources widely used in empirical research on digital ventures, with further details described in Appendix Sections B.1 and B.2.

Table \ref{tab:sumstats} reports descriptive statistics for the variables used in the venture-level analyses. Panel A presents statistics for the full sample. The outcome Survival 1 Year After Launch has a mean of 0.424. Panel B presents statistics for the subsample of ventures that survived for at least one year after launch. Within this subsample, the outcome Log (New Funding \$ Mil Within 1 Year) has a mean of 0.032.\footnote{Ventures in the one-year survival subsample have a 3.8\% likelihood of raising external venture financing within a year of launch, whereas this likelihood is close to zero among ventures that failed within a year of launch. Therefore, we restrict the data to the one-year survival sample for analyzing financing outcome.} In both samples, the share of founding teams with software experience is approximately 35\%. However, ventures in the one-year survival subsample tend to have founding teams with stronger observable credentials. Relative to the full sample, this subsample includes fewer solo founders and more serial founders, founders with prior employment in top startup ecosystems, founders with degrees from top engineering universities, female founders, and founders holding a master’s or higher degree. Finally, Panel C reports the correlations among the variables in the full sample.

\subsection{Empirical Methods} \label{subsec:empirical}

\subsubsection{Measuring Product Categories' Exposure to AI Coding} \label{subsubsec:measure_exposure}

To capture the variation in susceptibility to AI coding across ventures, we compute a product-category level measure of exposure to AI coding, defined as the extent to which a representative product in the category can be built using \textit{AI-assisted development where the developer describes what they want in natural language and iterates with an AI coding assistant without deep expertise in the technologies}.

The measurement involves a three-step text-based procedure. First, we recover the latent structure of the product space by clustering embeddings of product descriptions. The clustering corpus comprises more than 60,000 first-time venture launches on Product Hunt between 2019Q1 and 2022Q2, before the emergence of AI coding tools. We combine the tagline and description (which must be non-empty) of each product into a single text input. Using Qwen3-Embedding-4B, a leading open-weight pretrained text-embedding model,\footnote{We selected Qwen3-Embedding-4B because it performed best among leading open-weight alternatives including Qwen3-Embedding-8B, BGE-M3, BGE-Large, E5-Large, E5-Large-Instruct, and Nomic Embed. Fine-tuning did not yield performance improvements (see Appendix Section B.4 for further details.)} we represent each product as a 2,560-dimensional embedding vector. Following standard practice, we run agglomerative hierarchical clustering with average linkage and cosine distance on these embeddings, after using UMAP to reduce their dimensionality to 50. Maximizing the silhouette score across alternative cuts of the resulting hierarchy identifies 310 clusters as optimal. Appendix Figure A1 visualizes the hierarchical structure of these clusters labeled by frequent keywords. These clusters serve as distinct product categories, with each product assigned to exactly one of them.\footnote{Product Hunt’s self-reported categories are too coarse and imbalanced to reliably measure exposure to AI coding. For example, \textit{Productivity} alone accounts for 20\% of all launches.} 

Second, we measure each product category’s exposure to AI coding by scoring a random sample of its products.\footnote{The sampling pool is restricted to products launched between 2020Q1 and 2022Q2 that have a non-missing description, at least 19 words across the tagline and description, and a distance from the assigned cluster centroid below the 95th percentile. Every product category contains at least 10 eligible products.} We randomly sample 15 products per category, or all eligible products when fewer than 15 are available. Using the Claude API, we apply a detailed classification prompt developed with Claude Opus 4.5 (see Appendix Table A1 for the exact prompt, which specifies assessment criteria, examples, and calibration guidelines) and implemented with Claude Sonnet 4 at zero temperature. Each product receives a raw score from 0 to 4, as defined in Appendix Table A2. We calculate the aggregate score at the product-category level as the median score across sampled products.

Finally, we assign every venture in the full sample -- including those launched after 2022Q2 -- to the product category whose cluster centroid in the product space is closest to the text embedding vector of its first launch. We then convert each product category’s score into three groups: full exposure (4), partial exposure (2–3), and minimal exposure (0–1), as described in Appendix Table A2. Full exposure indicates that the median product’s core functionality follows standard patterns and can be produced by AI coding end-to-end. Partial exposure indicates that AI coding tools can produce some core components, while others require human expertise. Minimal exposure indicates that AI coding tools can produce little to none of the product’s core functionality. Appendix Figure A2 shows the distribution of category sizes in Panel (A) and quarterly venture launches across the three exposure groups in Panel (B). The descriptive patterns indicate that launches in the full- and partial-exposure groups increased substantially beginning in 2022Q3, whereas launches in the minimal-exposure group did not experience a comparable increase.

Table \ref{tab:vibecodability_examples} presents representative examples of ventures’ initial launches on Product Hunt across levels of AI coding exposure. For each raw score from 0 to 4, we show two products from the same product category, one launched before and one after mid-2022. These examples illustrate aspects of product development that may remain difficult to automate fully with AI coding tools, including but not limited to specialized knowledge in non-software domains (e.g., \textit{Shimmer ADHD} and \textit{Legion Health}), performance optimization (e.g., \textit{Supabase} and \textit{Releem}), hardware integration (e.g., \textit{Vagon} and \textit{LayerOps}), and safety-critical control systems (e.g., \textit{PowerX} and \textit{SolarPunk}).

\subsubsection{Estimation Strategy: Difference-in-Differences Regression Models} \label{subsubsec:did_analyses}

A pivotal milestone in the global diffusion of AI coding tools was GitHub Copilot’s transition from technical preview to general availability in June 2022. Around the same time, Replit introduced its ``Generate Code'' feature, which allowed users to describe a desired function in natural language and receive AI-generated code in response.\footnote{\url{https://replit.com/blog/generate-code}. Replit is particularly relevant because its large user base and browser-based development environment make AI-assisted coding widely accessible not only to professional developers, but also to digital entrepreneurs with non-technical backgrounds.} Prior research documents a sharp increase in AI-generated code in open-source repositories worldwide \citep{daniotti2026using}, consistent with rapid adoption of AI coding practices around this time. The AI coding platforms market subsequently expanded rapidly through the introduction of AI features by established platforms (e.g., Vercel and Bubble) and the rise to popularity of new entrants (e.g., Lovable). Appendix Table A3 presents a timeline of major launches of AI coding tools through the end of 2025.

Our identification strategy requires pinpointing when the availability of AI coding tools began to affect observable product launches by digital ventures. This effect should emerge with a delay because entrepreneurs engaging with AI coding tools need a non-trivial amount of time to develop a launch-ready product. As reported in Panel A of Table \ref{tab:sumstats}, the median interval between website creation and the first public launch is three months. We therefore define 2022Q4 -- approximately three months after the public release of GitHub Copilot and Replit’s ``Generate Code’’ feature -- as the beginning of the treatment period.
To test Hypotheses 1a and 1b, we estimate product-category level regressions described as follows. Equation \ref{eq:reg1} shows a difference-in-differences (DiD) regression specification for testing H1a, where the outcome $Ln(Entry_{ct})$ is the log number of new ventures launched in product category $c$ during year-quarter $t$. The underlying data is a balanced panel that includes category-quarter observations when no entry occurred, for which the variable equals zero. The regression controls for product-category fixed effects $\xi_{c}$ and year-quarter fixed effects $\eta_{t}$. Under the parallel-trends assumption, the coefficient $\alpha$ on $ExposureAny_{c}\times Post_{t}$ identifies difference between the change in entrepreneurial entry after 2022Q3 for product categories with $ExposureAny_{c}=1$ (in which the median product is either partially or fully exposed to AI coding, as clarified in Appendix Table A2) and the contemporaneous change for product categories with $ExposureAny_{c}=0$. 
\begin{equation} \label{eq:reg1}
Ln(Entry_{ct}) = \alpha ExposureAny_{c}\times Post_{t} + \xi_{c} + \eta_{t} + \epsilon_{ct}
\end{equation}

For testing H1b, we change the outcome variable in equation \ref{eq:reg1} into the share of ventures with at least one software-experienced founder by product category and year-quarter.

To test Hypotheses 2a and 2b, respectively, we estimate venture level regressions specified in equation \ref{eq:reg2} using the following outcome variables. Survival one year after launch is measured as an indicator equal to 1 if the venture’s website receives at least 50 monthly visits a year after its first launch and 0 otherwise.\footnote{We use 50 rather than zero visits as the threshold to distinguish meaningful operating activity from measurement noise, and our results are robust to alternative thresholds such as zero.} Log new funding raised within one year equals the natural logarithm of one plus the total external funding raised (in \$million) during the year after first launch (excluding pre-launch funding).\footnote{The variable equals zero for ventures that raise no funding during the one-year period after first launch.} $ExposureIntensity_{c}$ takes integer values from 0 to 2, and increases by one from minimal to partial exposure, and from partial to full exposure, respectively (see Appendix Table A2). Conditional on product-category fixed effects $\phi_{C(i)}$, launch-quarter fixed effects $\psi_{L(i)}$, and venture-level controls $X_i$, the coefficient $\beta$ on $ExposureIntensity_{C(i)}\times Post_{L(i)}$ identifies the effect of category-level exposure to AI coding on venture survival and financing outcomes.
\begin{equation} \label{eq:reg2}
Y_{i} = \beta ExposureIntensity_{C(i)}\times Post_{L(i)} + \gamma X_{i} + \phi_{C(i)}+\psi_{L(i)}+\nu_{i}
\end{equation}

For testing H2b and H3b, we add interaction terms to equation \ref{eq:reg2} between the key regressors and $SoftwareExperience_{i}$ which is a binary indicator for whether at least one founder was recently employed in a software engineering role outside the focal venture $i$.\footnote{This includes general software engineering as well as domain-specific roles such as web or mobile development, front-end or back-end development, system architect, or quality assurance. An employment episode is considered recent if it (1) lasted more than six months, (2) began at least one year before and (3) was ongoing or ended no more than two years before the venture’s launch.}

In all regression specifications, we cluster standard errors at the product-category level. Venture-level control variables include fixed effects for founding team size, website creation month, and founder age group, gender, education, and employment background, as well as the log number of same-quarter new ventures launched in the same product category.

\section{Analyses and Results} \label{sec:results}

\subsection{Results on Entrepreneurial Entry} \label{subsec:baseline}

\subsubsection{Exposure to AI Coding Increases Entrepreneurial Entry} \label{subsec:baseline_a}

\begin{figure}[htbp]
\centering
\caption{Estimated Effect of AI Coding on New Venture Launches}\label{fig:nobs}
    \includegraphics[width=0.85\textwidth]{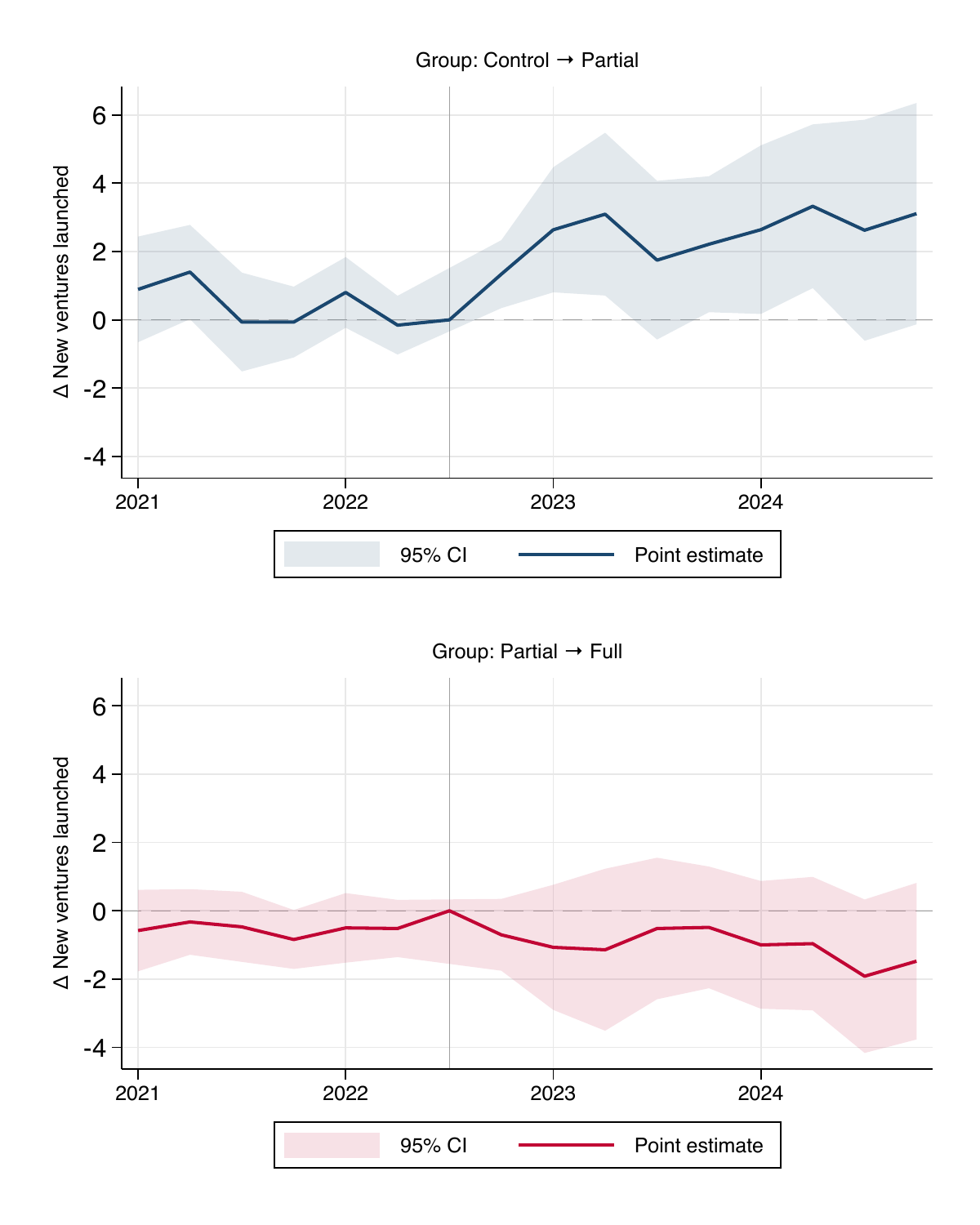}
\vspace{0.4em}
\begin{minipage}{0.85\textwidth}
\footnotesize\singlespacing
\noindent\textit{Note}: The lines and shaded areas display the the estimated differences in new venture launches and the 95\% confidence intervals around them, based on event study model results. In the top subfigure, the y-axis represents the estimated difference in new venture launches between partially and minimally exposed product categories. In the bottom subfigure, the y-axis represents the estimated difference in new venture launches between fully and partially exposed product categories. The vertical line marks 2022Q3, the omitted reference quarter.
\end{minipage}
\end{figure}

Table \ref{tab:nobs} reports the results for testing Hypothesis 1a. In Model 1, partial or full AI coding exposure increases the number of new ventures launched after treatment by 21\% ($p=0.002$). Model 2 replaces the post-treatment indicator with year indicators, using 2022 as the omitted reference year, and the results suggest somewhat larger effects in 2024 than in 2023. To further assess the parallel-trends assumption, Figure \ref{fig:nobs} reports event-study estimates of new venture entry by the number of quarters relative to treatment. The top panel compares partially exposed categories ($ExposureIntensity$=1) with minimally exposed categories ($
ExposureIntensity$=0), while the bottom panel compares fully exposed categories ($ExposureIntensity$=2) with partially exposed categories. In both panels, pre-treatment estimates are statistically indistinguishable from zero, consistent with the parallel-trends assumption, while the estimates are positive in the top panel and close to zero in the bottom panel in the post-treatment period. Overall, these results support H1a.

\subsubsection{Differential Entry by Founders’ Software Experience} \label{subsubsec:mechanism_software}
Table \ref{tab:software} reports the results for testing Hypothesis 1b. Model 1 shows that, in the post-treatment period, the share of ventures with software-experienced founders increases by 5.1 percentage points more (compared to the overall average of 34.3\%) in product categories with any exposure (i.e., either partial or full) than in minimally exposed categories ($p=0.038$). Model 2 builds on Model 1 by additionally controlling for the number of ventures launched in each product category and year-quarter. The estimated effect remains statistically significant and largely unchanged. These results imply that the quantity of entrepreneurial entry among founders without software experience into product categories exposed to AI coding increased less that among founders with prior software experience. Finally, Model 3 interacts $ExposureAny$ with year indicators, using 2022 as the omitted baseline year. The statistically insignificant coefficient for 2021 is consistent with the assumption of no pre-trends, while the coefficient for year 2024 is somewhat smaller than that for year 2023, suggesting that the selection of software experienced founders into entry in exposed categories declined over time. Overall, these results are consistent with H1b.

\subsection{Results on Venture Survival among Entrants} \label{subsec:outcome}
\subsubsection{Exposure to AI Coding Decreases Survival of Entrants}\label{subsubsec:outcome_survival}

Table \ref{tab:survival} reports the full-sample estimates for testing Hypothesis 2a. Model 1 shows that in the post-treatment period, the probability of ventures surviving one year after first launch suffered declines of 3.5\% ($p$=0.050) in partially exposed categories and a larger 5.8\% ($p$=0.003) in fully exposed categories. In Model 2, the treatment-effect estimates remain largely unchanged after adding as regressors the log number of ventures launched in the same product category and year-quarter to capture market-level competition, along with controls for venture age, founding-team size, and founders’ demographic, educational, and employment backgrounds. In Model 3, the regressors are replaced by $ExposureIntensity$ interacted with year indicators, with 2022 serving as the omitted baseline year. Consistent with the parallel trends assumption, the coefficient estimate for 2021 is small and statistically insignificant. Meanwhile, the coefficient estimates for 2023 and 2024 are negative and statistically significant at the 1\% level. These results provide broad support for H2a.

\subsubsection{Differential Effects on Venture Survival by Founders’ Software Experience}\label{subsubsec:moderate_survival}

Table \ref{tab:moderate_survival_software} reports the estimates for testing Hypothesis 2b. In Model 1, founders’ software experience completely offsets the negative baseline effect of partial exposure to AI coding, as shown by the coefficient of 6.9\% on the triple interaction among $ExposurePartial$, $Post$, and $SoftwareExperience$ ($p$=0.057). However, we do not find a commensurate moderating effect for $ExposureFull$.

The underlying samples of Models 2 and 3 are split into ventures with and without a software-experienced founder, respectively. As expected, survival rate does not suffer in partially exposed product categories among ventures with software-experienced founders, while ventures in fully exposed categories are still less likely to survive post-treatment (although the coefficient is not statistically significant). By contrast, ventures without software-experienced founders experience a decline in survival rate in both partially and fully exposed product categories. Overall, these findings support H2b.

\subsection{Results on Venture Financing among Survivors} \label{subsec:moderate}

\subsubsection{Exposure to AI Coding Increases Funding among Survivors}\label{subsubsec:outcome_funding}

Table \ref{tab:funding} reports the regression results for testing Hypothesis 3a. Because venture capital (VC) financing is rare even among technology ventures, the outcome is highly skewed toward zero with most ventures raising no funding. We therefore estimate the financing models among ventures that survived for at least one year, as this subsample includes nearly all ventures that raised external venture financing during that period.\footnote{Among ventures that survived for at least one year, 3.8\% raised venture financing within the first year after their first launch, whereas the corresponding share among ventures that failed within a year was close to zero.} Model 1 shows that in the post-treatment period, the amount of VC financing raised by a surviving venture within a year of launch increased by 3.6 percent ($p$=0.080) in partially exposed categories and a larger 6.8 percent ($p$=0.001) in fully exposed categories. The estimates from Model 2 show that the treatment-effect estimates remain largely the same, when the model additionally includes as regressors the log number of ventures launched in the same product category and year-quarter to capture market-level competition, along with controls for venture age, founding-team size, and founders’ demographic, educational, and employment backgrounds. In Model 3, the regressors are replaced by $ExposureIntensity$ interacted with year indicators, with 2022 serving as the omitted baseline year. Consistent with the parallel trends assumption, the coefficient estimate for 2021 is small and statistically insignificant. Meanwhile, the coefficient estimates for 2023 and 2024 are positive and statistically significant at the 5\% level. These results provide broad support for H3a.

\subsubsection{Differential Effects on Venture Funding by Founders’ Software Experience}\label{subsubsec:moderate_funding }

Table \ref{tab:moderate_funding_software} reports the regression results for testing Hypothesis 3b. In Model 1, the baseline effects of partial and full exposure to AI coding on venture financing amount are null, as indicated by insignificant and near-zero coefficients. Founders’ software experience positively moderates the effects of partial and full exposure on venture financing post-treatment with similar magnitudes, as shown by the coefficients of 9.5\% ($p$=0.017) on the interaction between $ExposurePartial$ and $Post\times SoftwareExperience$ and 9.8\% ($p$=0.013) on the interaction between $ExposureFull$ and $Post\times SoftwareExperience$. 

The underlying samples of Models 2 and 3 are split into ventures with and without software-experienced founders. For ventures with a software-experienced founder, both partial and full exposure are associated with large post-treatment increases of 11.0\% ($p$=0.003) and 14.5\% ($p$<0.001) respectively in venture financing raised within one year of launch. By contrast, exposure to AI coding has no effect on venture financing when no founder has prior software experience. Overall, these findings support H3b.

\subsection{Mechanisms and Robustness Tests} \label{subsec:mechanism}

\subsubsection{Ruling Out Selection of Lower-Quality Ventures into Exposed Categories}\label{subsubsec:mechanism_quality}

For Hypothesis 2a and 2b, we rule out an alternative explanation that lower-quality entrepreneurs disproportionately enter more exposed categories. Appendix Table A4 provides evidence against this alternative explanation and indicates no adverse selection of lower-quality entrepreneurs into exposed product categories. Model 1 shows no differential change in the share of solo-founded ventures relative to team-founded ventures in exposed categories after the treatment, while Model 2 shows no differential change in the share of ventures founded by serial entrepreneurs relative to those founded by first-time entrepreneurs. 

Models 3 and 4 focus on founders’ access to high-quality networks through prior employment in leading startup ecosystems or education at top engineering universities. The results run counter to the adverse-selection explanation. Model 3 shows that, in exposed categories after treatment, the share of ventures founded by individuals with recent employment experience in startup ecosystems\footnote{These are Density Leaders ranked by Dealroom’s Global Tech Ecosystem Index 2026.} increased by 4 percentage points relative to a baseline mean of 32\% ($p=0.098$). Model 4 shows that the share of ventures founded by individuals with at least a bachelor’s degree from a top engineering university\footnote{These are Top-20 universities in Engineering and Technology according to 2026 QS World University Rankings. MIT, Stanford, ETH Zurich, Oxford, and Cambridge are the top five universities on this list.} increased by 3 percentage points relative to a baseline mean of 5.5\% ($p=0.011$). Rather than disproportionately expanding entry among less-resourced founders, exposure to AI coding appears to have generated a larger increase in entry among graduates of top engineering universities and, to a lesser extent, former employees in leading startup ecosystems.

Models 5--7 focus on founders’ demographic and educational characteristics and again provide little evidence of adverse selection. Model 5 shows that the share of ventures with female founders declined by 3.6 percentage points in exposed product categories after treatment, relative to a baseline mean of 17\%, although the estimate is only marginally significant ($p=0.061$). Models 6 and 7 show no differential change in founders’ age or in the share of ventures founded by individuals with advanced degrees (i.e., a master’s degree or higher). 

Taken together, instead of suggesting that entrants become lower quality, the evidence indicates that they are no less likely to possess the technical skills needed for product development or the business skills needed to launch and operate a new venture post-treatment.

\subsubsection{Ruling Out Educational and Local Networks as Alternative Explanations}\label{subsubsec:mechanism_network}
Frontier AI labs and advances in deep learning are concentrated in a small number of locations and depend on scarce top talent trained at a handful of leading engineering institutions. Therefore, a potential alternative explanation to H2b and H3b is that they are driven by founders’ access to external resources through university alumni and former coworkers rather than software experience itself. These networks may help founders recruit higher-quality co-founders and early engineers or obtain introductions to higher-quality potential investors. This explanation is plausible because the results in Section \ref{subsubsec:mechanism_quality} and Appendix Table A4 show that ventures founded by graduates of top engineering universities and, to a lesser extent, former employees in major startup ecosystems accounted for a larger share of post-treatment entry into categories exposed to AI coding. 

Appendix Table A5 reports the results from one test of this alternative explanation, which re-estimates the models in Table \ref{tab:software} after excluding ventures whose founders graduated from a top engineering university. Across all models, the results are highly similar to the original estimates obtained from the full sample.

Appendix Table A6 reports the results from another test replacing $SoftwareExperience$ with $HubEmployment$ and $TopEngineeringDegree$ as moderators in the regressions underlying Tables \ref{tab:moderate_survival_software} and \ref{tab:moderate_funding_software}. All relevant coefficients on the triple interactions are small (or in the opposite direction) and statistically insignificant, indicating that neither educational nor professional networks account for the complementary capabilities associated with software experience in improving post-treatment venture survival or facilitating external financing in exposed categories.

\subsubsection{Product Innovation Mechanism for the Venture Financing Results}\label{subsubsec:mechanism_aiproduct}

A key mechanism underlying H3a and H3b is that ventures in product categories more exposed to AI coding may be more likely to embed AI capabilities in their core offerings, and attract greater financing as a result. We test this mechanism using a continuous measure, ranging from 0 to 1, of the likelihood that a venture’s product description reflects an AI-native offering. We use this measure as an intermediate outcome to assess whether AI-native product innovation helps explain the financing results reported in Table \ref{tab:moderate_funding_software}. Appendix Table A7 presents the scoring prompt in panel (a) and the distribution of scores generated by parsing the product descriptions through the Claude Sonnet 5 API in panel (b).

Appendix Table A8 reports the estimation results on models analogous to those in Table \ref{tab:moderate_funding_software}, replacing external financing with AI-native likelihood as the outcome variable. The results are highly consistent with the proposed mechanism underlying H3a and H3b: Exposure to AI coding increases the AI-native likelihood of ventures’ product offerings in exposed categories, but only for ventures with a software-experienced founder, and not for ventures without a software-experienced founder.

\section{Discussion} \label{sec:discuss}

Prior research shows that access to digital infrastructural resources lead to an increasing quantity and variety of new products \citep{waldfogel2017digitization} and the democratization of entrepreneurial entry \citep{kerr2014entrepreneurship,stroube2025mapping}. Yet it is not clear when these technical resources may enable viable ventures that create lasting value, and what configuration of organizational capabilities facilitate innovation from these technical resources. More recently, the rise of AI coding tools plays a similar role in lowering barriers to entrepreneurial entry, but its usage intertwines technical and operational sides of the business, rather than letting entrepreneurs largely outsource the technical layer as they can with cloud computing or low-code tools.

To examine the effects of AI coding tools on entrepreneurial entry, survival, and financing, we combine data on digital ventures and founder career histories, and leverage the intrinsic differences among product categories’ exposure to AI coding for econometric identification. Our findings show a significantly increase in first-time launches of new ventures, which is larger among founders with software experience, in product categories partially and fully exposed to AI coding. One year from the first launch, new ventures in product categories more exposed to AI coding experience lower likelihood of survival, but those that survived raise more financing on average. Founders’ software experience has a positive moderating effect on venture survival in partially but not fully exposed product categories, and explains virtually all of the increase in external financing raised within a year of first launch. 

Our study contributes to research on how access to digital technologies can spur entrepreneurial entry across wider populations \citep{kerr2014entrepreneurship,fossen2021digitalization,dushnitsky2021low,stroube2025mapping}. Our analyses extend this research by analyzing performance outcomes after venture creation, showing that widespread availability of AI coding tools decreases venture survival and increases external financing in exposed product markets. These findings offer evidence for effects of AI beyond resource substitution for software tasks, in the potential to reorganize product development processes and innovate business models by making AI the core of the venture’s value proposition. They resonate with studies on the pervasive impact of digitalization \citep{waldfogel2017digitization} which not only raised the quantity and diversity of products \citep{benner2023changing} but also led to the emergence of digital-native products created in digital form from their inception \citep{benner2016song}. 

This research also adds to the growing literature on how technological change reshapes the role of human expertise in contributing to firm productivity and innovation \citep{choudhury2020machine,krakowski2023artificial,devigili2026skill,sauermann2026extension,cavalli2026ai}. The individual mechanism that organizational members with greater expertise are more productive in leveraging AI \citep{jia2024and,otis2026uneven} aggregates to the firm level for entrepreneurial ventures. We find that founders’ software expertise complement AI coding tools in shaping survival and financing outcomes, aligned with emerging evidence from recent studies across different settings \citep{bena2026prompted,kim2026ai}. Our results are consistent with labor economics research showing that the automation of generic tasks can increase the value of expertise \citep{autor2025expertise}, and that firms may find it economically rational to automate some but not all tasks even in the longer run \citep{li2026economics}. While AI coding tools may not render software expertise obsolete, it may reshape engineer’s work by redistributing their efforts from manual coding to performing high-level activities such as architectural design, quality assurance, and performance optimization.

From a practical and managerial standpoint, clearly many of the new digital products created in categories exposed to AI coding may not enable the creation of products with durable value or lead to a lasting viable business. Overall, new ventures with scarce complementary capabilities may still turn a profit in areas where product development requires expertise that cannot be automated by AI. On the other hand, AI seems to catalyze product innovations by which ventures, especially those founded by software-experienced individuals, achieve unprecedented speed of scaling that threaten to replace traditional software business models.

\subsection{Limitations and Future Research}
Data limitation prevents us from observing venture-level adoption and usage of specific AI coding tools. Hence, we are unable to distinguish the use of AI coding tools from that of conversational AI tools (e.g., ChatGPT) and AI image-generation tools (e.g., Midjourney) at the venture level. To avoid this drawback, our empirical approach uses inherent differences across product categories in the extent to which product development can take place via AI coding practices, while assuming that founders have broadly similar access to other forms of AI. The identifying assumption is that any effects of using AI for other activities such as ideation or marketing must arise similarly across the treatment and control groups.

Another limitation is that the treatment captures the early wave of AI coding practices beginning in mid-2022. AI capabilities have  advanced rapidly with the rise of agentic systems since late 2025, raising questions about whether our findings will generalize to an era in which longer-horizon agents increasingly supplant earlier generations of AI-assistant tools. However, there are reasons to expect our core findings to remain directionally relevant. IT professionals have historically responded to the obsolescence of existing skills by acquiring new ones and keeping pace with frontier knowledge in a rapidly evolving sector \citep{bapna2013human,horton2025death}. Agentic AI may transform rather than eliminate the tasks through which software expertise creates value, in areas such as architectural judgment, code maintenance, and domain-specific system design. Our analysis does not, however, consider organizational capabilities enabled by agentic systems, such as multi-agent coordination or the use of one agent to verify another’s work. Examining how such capabilities reshape organizational design goes beyond the scope of this study but represents a promising direction for future research.

\pagebreak
\singlespacing
\bibliographystyle{apalike}
\bibliography{references}

\singlespacing

\pagebreak

\clearpage

\begin{table}[htbp]
\centering
\scriptsize
\def\sym#1{\ifmmode^{#1}\else\(^{#1}\)\fi}
\caption{Descriptive Statistics and Pairwise Correlations.}
\label{tab:sumstats}

\begin{tabular}{lcccccc}
\toprule
\multicolumn{7}{l}{\textbf{Panel (A): Full Sample}} \\
\midrule
                    & Mean & Std.\ Dev. & P10 & P50 & P90 & N \\
\midrule
Survival 1 Year After Launch & 0.424 & 0.494 & 0 & 0 & 1 & 33155 \\
Log (New Funding \textdollar{} Mil Within 1 Year) & 0.015 & 0.160 & 0 & 0 & 0 & 33155 \\
Months from Website to First Launch & 7.991 & 11.530 & 0 & 3 & 26 & 33155 \\
Software Experience & 0.354 & 0.478 & 0 & 0 & 1 & 33155 \\
Solo Founder & 0.792 & 0.406 & 0 & 1 & 1 & 33155 \\
Serial Founder & 0.369 & 0.483 & 0 & 0 & 1 & 33155 \\
Prior Employment in Hub & 0.327 & 0.469 & 0 & 0 & 1 & 33155 \\
Top Engineering Education & 0.057 & 0.231 & 0 & 0 & 0 & 33155 \\
Has Female Founder & 0.160 & 0.367 & 0 & 0 & 1 & 33155 \\
Median Founder Age & 30.086 & 6.853 & 22 & 29 & 40 & 33155 \\
Has Master's Degree & 0.364 & 0.481 & 0 & 0 & 1 & 33155 \\
\bottomrule
\end{tabular}

\vspace{0.6em}

\begin{tabular}{lcccccc}
\toprule
\multicolumn{7}{l}{\textbf{Panel (B): 1-Year Survival Ventures Sample}} \\
\midrule
                    & Mean & Std.\ Dev. & P10 & P50 & P90 & N \\
\midrule
Log (New Funding \textdollar{} Mil Within 1 Year) & 0.032 & 0.240 & 0 & 0 & 0 & 14042 \\
Months from Website to First Launch & 9.845 & 11.936 & 0 & 5 & 29 & 14042 \\
Software Experience & 0.355 & 0.479 & 0 & 0 & 1 & 14042 \\
Solo Founder & 0.734 & 0.442 & 0 & 1 & 1 & 14042 \\
Serial Founder & 0.415 & 0.493 & 0 & 0 & 1 & 14042 \\
Prior Employment in Hub & 0.352 & 0.477 & 0 & 0 & 1 & 14042 \\
Top Engineering Education & 0.064 & 0.245 & 0 & 0 & 0 & 14042 \\
Has Female Founder & 0.182 & 0.386 & 0 & 0 & 1 & 14042 \\
Median Founder Age & 30.239 & 6.553 & 22 & 30 & 40 & 14042 \\
Has Master's Degree & 0.387 & 0.487 & 0 & 0 & 1 & 14042 \\
\bottomrule
\end{tabular}

\vspace{0.6em}

{\scriptsize
\setlength{\tabcolsep}{2.1pt}
\renewcommand{\arraystretch}{0.88}
\textbf{Panel (C): Pairwise Correlations}\par\vspace{0.25em}
\resizebox{\textwidth}{!}{%
\begin{tabular}{l*{11}{c}}
\toprule
          & (1) & (2) & (3) & (4) & (5) & (6) & (7) & (8) & (9) & (10) & (11) \\
\midrule
(1) Survival 1 Year After Launch & 1.00 &  &  &  &  &  &  &  &  &  &  \\
(2) Log (New Funding \textdollar{} Mil Within 1 Year) & 0.09 & 1.00 &  &  &  &  &  &  &  &  &  \\
(3) Months from Website to First Launch & 0.14 & 0.04 & 1.00 &  &  &  &  &  &  &  &  \\
(4) Software Experience & 0.00 & 0.01 & -0.03 & 1.00 &  &  &  &  &  &  &  \\
(5) Solo Founder & -0.12 & -0.11 & -0.05 & -0.15 & 1.00 &  &  &  &  &  &  \\
(6) Serial Founder & 0.08 & 0.03 & 0.02 & -0.07 & -0.24 & 1.00 &  &  &  &  &  \\
(7) Prior Employment in Hub & 0.04 & 0.07 & -0.00 & 0.03 & -0.20 & 0.12 & 1.00 &  &  &  &  \\
(8) Top Engineering Education & 0.03 & 0.07 & 0.00 & 0.01 & -0.14 & 0.04 & 0.19 & 1.00 &  &  &  \\
(9) Has Female Founder & 0.05 & 0.02 & 0.03 & -0.02 & -0.25 & 0.04 & 0.06 & 0.04 & 1.00 &  &  \\
(10) Median Founder Age & 0.02 & 0.01 & 0.07 & -0.10 & 0.04 & 0.06 & 0.10 & -0.00 & -0.01 & 1.00 &  \\
(11) Has Master's Degree & 0.04 & 0.05 & 0.03 & -0.03 & -0.21 & 0.04 & 0.13 & 0.18 & 0.11 & 0.17 & 1.00 \\
\bottomrule
\end{tabular}%
}
}

\vspace{0.35em}

\begin{minipage}{0.98\textwidth}
\singlespacing\scriptsize
\textit{Note}: Panel (A) reports descriptive statistics for the full sample. Panel (B) reports descriptive statistics for ventures that survive one year after launch. Panel (C) reports Pearson correlation coefficients for each pair of variables in the full sample.
\end{minipage}
\end{table}

\begin{landscape} \begin{center} \scriptsize \setlength{\tabcolsep}{3pt} \renewcommand{\arraystretch}{1.15} \begin{longtable}{ >{\RaggedRight\arraybackslash}p{2.2cm} >{\centering\arraybackslash}p{1.2cm} >{\centering\arraybackslash}p{1.1cm} >{\centering\arraybackslash}p{1.2cm} >{\RaggedRight\arraybackslash}p{4.3cm} >{\RaggedRight\arraybackslash}p{10cm} >{\centering\arraybackslash}p{1.6cm} } \caption{Examples by AI Coding Exposure Classification} \label{tab:vibecodability_examples}\\ \toprule \shortstack{Company} & \shortstack{Category\\ID} & \shortstack{Score\\(0--4)} & \shortstack{Exposure\\(0--2)} & \shortstack{Product Hunt Tagline} & \shortstack{Product Hunt Description} & \shortstack{Launch Date} \\ \midrule \endfirsthead \multicolumn{7}{l}{\textit{Table \thetable{} continued}}\\ \toprule \shortstack{Company} & \shortstack{Category\\ID} & \shortstack{Score\\(0--4)} & \shortstack{Exposure\\(0--2)} & \shortstack{Product Hunt Tagline} & \shortstack{Product Hunt Description} & \shortstack{Launch Date} \\ \midrule \endhead \midrule \multicolumn{7}{r}{\textit{Continued on next page}}\\ \endfoot \bottomrule \endlastfoot \rowcolor{vcDarkGreen!15} Routine & 96 & 4 & 2 & Tasks, notes and calendar into one fast productivity app & Routine (YC W21) combines tasks, notes and calendar into one fast productivity app. Capture tasks and notes in one central place; secure time to work on what is most important; and link information to time, such as meeting notes. Give it a try at routine.co. & Apr 2021 \\ \rowcolor{vcDarkGreen!15} Hoop & 96 & 4 & 2 & AI task management for busy professionals & Hoop connects to tools such as Google Meet, Zoom, and Slack and uses AI to capture all your tasks. It provides one consolidated list that follows you from meeting to Slack channel to meeting. & Sep 2024 \\ \rowcolor{vcDarkBrownYellow!15} Shimmer ADHD & 113 & 3 & 1 & Find people who get you and improve your mental well-being & Shimmer offers online video support groups for personal growth led by coaches. Members join a small group with shared identities and experiences, meet weekly through video chat, attend community events, and work toward becoming the best version of themselves. & Jul 2021 \\ \rowcolor{vcDarkBrownYellow!15} Legion Health & 113 & 3 & 1 & World-class, AI-enabled psychiatry covered by insurance & Legion Health is a high-quality, AI-enabled psychiatry network where patients can receive care from a provider who accepts their insurance. It uses large language models and AI agents to improve care quality, streamline operations, enhance patient communications, and perform other functions. & Oct 2024 \\ \rowcolor{vcYellowBrown!15} Supabase & 146 & 2 & 1 & The open source Firebase alternative & Supabase is an open-source backend-as-a-service that provides real-time databases, authentication, and API services. It enables developers to build and scale applications quickly without managing server infrastructure. & May 2020 \\ \rowcolor{vcYellowBrown!15} Releem & 146 & 2 & 1 & Database performance management tool for developers & Releem makes managing and tuning database servers easy. Its features include automatic performance tuning, rapid identification of slow queries, integrated monitoring and tuning, support for MySQL, MariaDB, and AWS RDS, and an open-source agent. & Jun 2024 \\ \rowcolor{vcLightRed!15} Vagon & 156 & 1 & 0 & Your personal high-performance computer on the cloud & Vagon is a personal high-performance computer in the cloud for creative users. It allows users to run graphics software such as Adobe or Autodesk products from any device and location, thereby providing flexible computing performance. & Mar 2020 \\ \rowcolor{vcLightRed!15} LayerOps & 156 & 1 & 0 & A centralized platform for multi- and hybrid-cloud management & LayerOps offers a centralized platform for multi-cloud and hybrid-cloud management. It enables applications to be deployed and managed across multiple public, private, and hybrid clouds without requiring users to master the technical characteristics of each provider. & Jun 2024 \\ \rowcolor{vcDarkRed!15} PowerX & 85 & 0 & 0 & Save \$400 per year on your water heating bill & PowerX makes water heating more efficient. Described as the Nest for water heating, it learns users' preferred water temperatures and heats only as much water as needed. PowerX claims average annual savings of \$400 and CO$_2$ reductions equivalent to the absorption of ten fully grown trees. & Feb 2020 \\ \rowcolor{vcDarkRed!15} SolarPunk & 85 & 0 & 0 & Stripe Atlas for solar deployments & Like Stripe Atlas for solar, SolarPunk reduces bureaucracy and paperwork to help the next generation of farmers start agrivoltaics projects. The dual use of land for farming under solar panels can increase yields by up to 160\% and create a healthier microclimate. & Nov 2024 \\ \end{longtable} \end{center} 
\vspace{-1.4cm}
\begin{minipage}{22.8cm}
\singlespacing\scriptsize \textit{Note}: For every raw AI coding exposure classification score from 0 to 4, the table displays a pair of Product Hunt first-time launch examples in the same product category with the score. For each pair of examples, one launch took place pre-treatment (before 2022) and the other took place post-treatment (in 2024).
\end{minipage}
\end{landscape}

\begin{table}[htbp]\centering
\small
\def\sym#1{\ifmmode^{#1}\else\(^{#1}\)\fi}
\caption{Treatment Effects on Log Number of New Ventures Launched\label{tab:nobs}}
\begin{tabular}{l*{2}{c}}
\toprule
                &\multicolumn{2}{c}{Log (First-Time Launches)}\\\cmidrule(lr){2-3}
                &\multicolumn{1}{c}{(1)}         &\multicolumn{1}{c}{(2)}         \\
\midrule
Exposure Any $\times$ Post&    0.205\sym{***}&                  \\
                &  (0.067)         &                  \\
\addlinespace
Exposure Any $\times$ (Year=2021)&                  &   -0.017         \\
                &                  &  (0.063)         \\
\addlinespace
Exposure Any $\times$ (Year=2023)&                  &    0.169\sym{***}\\
                &                  &  (0.064)         \\
\addlinespace
Exposure Any $\times$ (Year=2024)&                  &    0.219\sym{***}\\
                &                  &  (0.080)         \\
Product Category FEs&      Yes         &      Yes         \\
Launch YQ FEs   &      Yes         &      Yes         \\
\midrule
Avg. Outcome    &    1.833         &    1.833         \\
Observations    &    4,960         &    4,960         \\
\(R^2\)         &    0.721         &    0.721         \\
\bottomrule
\end{tabular}
\vspace{0.4em}
\begin{minipage}{0.98\textwidth}
\footnotesize\singlespacing
\noindent\textit{Note}: Robust standard errors in parentheses are clustered at the product-category level, and the p-values are  $^{*}p<0.10$, $^{**}p<0.05$, $^{***}p<0.01$. Columns 1-2 report estimates from product-category-by-quarter regressions in which the outcome is the natural logarithm of the number of new ventures launched. All specifications include product category and year-quarter fixed effects.
\end{minipage}
\end{table}

\begin{table}[htbp]\centering
\small
\caption{Treatment Effects on Software-Experienced Founders Share}\label{tab:software}
\begin{tabular}{l*{4}{>{\centering\arraybackslash}p{2cm}}}
\toprule
                &\multicolumn{3}{c}{Software-Experienced Founders Share} \\\cmidrule(lr){2-4}
                &\multicolumn{1}{c}{(1)}         &\multicolumn{1}{c}{(2)}         &\multicolumn{1}{c}{(3)}         \\
\midrule
Exposure Any $\times$ Post&    0.051\sym{**} &    0.052\sym{**} &                  \\
                &  (0.025)         &  (0.025)         &                  \\
\addlinespace
Exposure Any $\times$ (Year=2021)&                  &                  &    0.044         \\
                &                  &                  &  (0.031)         \\
\addlinespace
Exposure Any $\times$ (Year=2023)&                  &                  &    0.081\sym{**} \\
                &                  &                  &  (0.037)         \\
\addlinespace
Exposure Any $\times$ (Year=2024)&                  &                  &    0.059\sym{*}  \\
                &                  &                  &  (0.035)         \\
\addlinespace
Log (First-Time Launches)&                  &   -0.004         &   -0.004         \\
                &                  &  (0.011)         &  (0.011)         \\
Product Category FEs&      Yes         &      Yes         &      Yes         \\
Launch YQ FEs   &      Yes         &      Yes         &      Yes         \\
\midrule
Avg. Outcome    &    0.343         &    0.343         &    0.343         \\
Observations    &    4,682         &    4,682         &    4,682         \\
\(R^2\)         &    0.184         &    0.184         &    0.184         \\
\bottomrule
\end{tabular}
\vspace{0.4em}
\begin{minipage}{0.98\textwidth}
\footnotesize\singlespacing
\noindent\textit{Note}: Robust standard errors in parentheses are clustered at the product-category level, and the p-values are  $^{*}p<0.10$, $^{**}p<0.05$, $^{***}p<0.01$. Columns 1-3 report estimates from product-category-by-quarter regressions in which the outcome is the share of new ventures launched where at least one founder previously worked in a software role. All specifications include product category and year-quarter fixed effects.
\end{minipage}
\end{table}

\begin{table}[htbp]\centering
\small
\def\sym#1{\ifmmode^{#1}\else\(^{#1}\)\fi}
\caption{Treatment Effects on Venture Survival After Launch\label{tab:survival}}
\begin{tabular}{l*{3}{c}}
\toprule
                &\multicolumn{3}{c}{Survival 1 Year After Launch}        \\\cmidrule(lr){2-4}
                &\multicolumn{1}{c}{(1)}         &\multicolumn{1}{c}{(2)}         &\multicolumn{1}{c}{(3)}         \\
\midrule
Expo. Partial $\times$ Post&   -0.035\sym{*}  &   -0.031         &                  \\
                &  (0.018)         &  (0.019)         &                  \\
\addlinespace
Expo. Full $\times$ Post&   -0.058\sym{***}&   -0.051\sym{**} &                  \\
                &  (0.019)         &  (0.020)         &                  \\
\addlinespace
Expo. Intensity $\times$ (Year=2021)&                  &                  &   -0.017         \\
                &                  &                  &  (0.014)         \\
\addlinespace
Expo. Intensity $\times$ (Year=2023)&                  &                  &   -0.033\sym{***}\\
                &                  &                  &  (0.012)         \\
\addlinespace
Expo. Intensity $\times$ (Year=2024)&                  &                  &   -0.034\sym{***}\\
                &                  &                  &  (0.012)         \\
Venture Controls&       No         &      Yes         &      Yes         \\
Product Category FEs&      Yes         &      Yes         &      Yes         \\
Launch YQ FEs   &      Yes         &      Yes         &      Yes         \\
\midrule
Avg. Outcome    &    0.424         &    0.424         &    0.424         \\
Observations    &   33,155         &   33,155         &   33,155         \\
\(R^2\)         &    0.039         &    0.084         &    0.084         \\
\bottomrule
\end{tabular}
\vspace{0.4em}
\begin{minipage}{16cm}
\footnotesize\singlespacing
\noindent\textit{Note}: Robust standard errors in parentheses are clustered at the product-category level, and the p-values are $^{*}p<0.10$, $^{**}p<0.05$, and $^{***}p<0.01$. Columns 1-3 report estimates from venture-level regressions in which the outcome is an indicator for whether the venture survived one year after first launch. Venture Controls include fixed effects for founding team size, website creation month, and founder age group, gender, education, and employment background, and the log number of same-quarter new ventures launched in the same product category. All specifications include product category and year-quarter fixed effects.
\end{minipage}
\end{table}

\begin{table}[htbp]\centering
\def\sym#1{\ifmmode^{#1}\else\(^{#1}\)\fi}
\caption{Software Experience Moderating Effects on Venture Survival}\label{tab:moderate_survival_software}
\small
{
\begin{tabular}{l*{3}{c}}
\toprule
                &\multicolumn{3}{c}{Survival 1 Year After Launch}        \\\cmidrule(lr){2-4}
Founder Experience:         &\multicolumn{1}{c}{All}&\multicolumn{1}{c}{Software=1}&\multicolumn{1}{c}{Software=0}\\
                &\multicolumn{1}{c}{(1)}&\multicolumn{1}{c}{(2)}&\multicolumn{1}{c}{(3)}\\
\midrule
Expo. Partial $\times$ Post&   -0.054\sym{***}&    0.007         &   -0.051\sym{**} \\
                &  (0.020)         &  (0.033)         &  (0.020)         \\
\addlinespace
Expo. Full $\times$ Post&   -0.057\sym{***}&   -0.048         &   -0.057\sym{***}\\
                &  (0.022)         &  (0.035)         &  (0.021)         \\
\addlinespace
Sftw. Expr.     &    0.007         &                  &                  \\
                &  (0.024)         &                  &                  \\
\addlinespace
Post $\times$ Sftw. Expr.&   -0.042         &                  &                  \\
                &  (0.032)         &                  &                  \\
\addlinespace
Expo. Partial $\times$ Sftw. Expr.&   -0.055\sym{**} &                  &                  \\
                &  (0.028)         &                  &                  \\
\addlinespace
Expo. Full $\times$ Sftw. Expr.&    0.011         &                  &                  \\
                &  (0.029)         &                  &                  \\
\addlinespace
Expo. Partial $\times$ Post $\times$ Sftw. Expr.&    0.069\sym{*}  &                  &                  \\
                &  (0.036)         &                  &                  \\
\addlinespace
Expo. Full $\times$ Post $\times$ Sftw. Expr.&    0.011         &                  &                  \\
                &  (0.037)         &                  &                  \\
Venture Controls&      Yes         &      Yes         &      Yes         \\
Product Category FEs&      Yes         &      Yes         &      Yes         \\
Launch YQ FEs   &      Yes         &      Yes         &      Yes         \\
\midrule
Avg. Outcome    &    0.424         &    0.424         &    0.423         \\
Observations    &   33,155         &   11,743         &   21,403         \\
\(R^2\)         &    0.084         &    0.121         &    0.085         \\
\bottomrule
\end{tabular}
}
\vspace{0.4em}
\begin{minipage}{16cm}
\footnotesize\singlespacing
\textit{Note}: Robust standard errors in parentheses are clustered at the product-category level, and the p-values are $^{*}p<0.10$, $^{**}p<0.05$, and $^{***}p<0.01$. Columns 1-3 report estimates from venture-level regressions in which the outcome is an indicator for whether the venture survived one year after first launch. All specifications include product category and year-quarter fixed effects, as well as venture controls, including fixed effects for founding team size, website creation month, and founder age group, gender, education, and employment background, and the log number of same-quarter new ventures launched in the same product category.
\end{minipage}
\end{table}

\begin{table}[htbp]\centering
\small
\def\sym#1{\ifmmode^{#1}\else\(^{#1}\)\fi}
\caption{Treatment Effects on Post-Launch Venture Financing\label{tab:funding}}
\begin{tabular}{l*{3}{c}}
\toprule
                &\multicolumn{3}{c}{Log (Funding \$ Mil Within 1 Year)} \\\cmidrule(lr){2-4}
                &\multicolumn{1}{c}{(1)}         &\multicolumn{1}{c}{(2)}         &\multicolumn{1}{c}{(3)}         \\
\midrule
Expo. Partial $\times$ Post&    0.036\sym{*}  &    0.033\sym{*}  &                  \\
                &  (0.021)         &  (0.020)         &                  \\
\addlinespace
Expo. Full $\times$ Post&    0.068\sym{***}&    0.063\sym{***}&                  \\
                &  (0.020)         &  (0.019)         &                  \\
\addlinespace
Expo. Intensity $\times$ (Year=2021)&                  &                  &   -0.011         \\
                &                  &                  &  (0.016)         \\
\addlinespace
Expo. Intensity $\times$ (Year=2023)&                  &                  &    0.021\sym{**} \\
                &                  &                  &  (0.010)         \\
\addlinespace
Expo. Intensity $\times$ (Year=2024)&                  &                  &    0.024\sym{**} \\
                &                  &                  &  (0.010)         \\
Venture Controls&       No         &      Yes         &      Yes         \\
Product Category FEs&      Yes         &      Yes         &      Yes         \\
Launch YQ FEs   &      Yes         &      Yes         &      Yes         \\
\midrule
Avg. Outcome    &    0.032         &    0.032         &    0.032         \\
Observations    &   14,041         &   14,041         &   14,041         \\
\(R^2\)         &    0.054         &    0.098         &    0.098         \\
\bottomrule
\end{tabular}
\vspace{0.4em}
\begin{minipage}{16cm}
\footnotesize\singlespacing
\noindent\textit{Note}: Robust standard errors in parentheses are clustered at the product-category level, and the p-values are $^{*}p<0.10$, $^{**}p<0.05$, and $^{***}p<0.01$. Columns 1-3 report estimates from venture-level regressions in which the outcome is log venture capital financing raised within one year after first launch. The sample includes ventures surviving one year after first launch. Venture Controls include fixed effects for founding team size, website creation month, and founder age group, gender, education, and employment background, and the log number of same-quarter new ventures launched in the same product category. All specifications include product category and year-quarter fixed effects.
\end{minipage}
\end{table}

\begin{table}[htbp]\centering
\def\sym#1{\ifmmode^{#1}\else\(^{#1}\)\fi}
\caption{Software Experience Moderating Effects on Venture Financing}\label{tab:moderate_funding_software}
\small
{
\begin{tabular}{l*{3}{c}}
\toprule
                &\multicolumn{3}{c}{Log (Funding \textdollar{} Mil Within 1 Year)}\\\cmidrule(lr){2-4}
Founder Experience:         &\multicolumn{1}{c}{All}&\multicolumn{1}{c}{Software=1}&\multicolumn{1}{c}{Software=0}\\
                &\multicolumn{1}{c}{(1)}&\multicolumn{1}{c}{(2)}&\multicolumn{1}{c}{(3)}\\
\midrule
Expo. Partial $\times$ Post&    0.000         &    0.110\sym{***}&   -0.006         \\
                &  (0.019)         &  (0.037)         &  (0.020)         \\
\addlinespace
Expo. Full $\times$ Post&    0.029         &    0.145\sym{***}&    0.022         \\
                &  (0.018)         &  (0.038)         &  (0.019)         \\
\addlinespace
Sftw. Expr.     &    0.084\sym{**} &                  &                  \\
                &  (0.034)         &                  &                  \\
\addlinespace
Post $\times$ Sftw. Expr.&   -0.097\sym{***}&                  &                  \\
                &  (0.036)         &                  &                  \\
\addlinespace
Expo. Partial $\times$ Sftw. Expr.&   -0.092\sym{**} &                  &                  \\
                &  (0.038)         &                  &                  \\
\addlinespace
Expo. Full $\times$ Sftw. Expr.&   -0.087\sym{**} &                  &                  \\
                &  (0.037)         &                  &                  \\
\addlinespace
Expo. Partial $\times$ Post $\times$ Sftw. Expr.&    0.095\sym{**} &                  &                  \\
                &  (0.040)         &                  &                  \\
\addlinespace
Expo. Full $\times$ Post $\times$ Sftw. Expr.&    0.098\sym{**} &                  &                  \\
                &  (0.039)         &                  &                  \\
Venture Controls&      Yes         &      Yes         &      Yes         \\
Product Category FEs&      Yes         &      Yes         &      Yes         \\
Launch YQ FEs   &      Yes         &      Yes         &      Yes         \\
\midrule
Avg. Outcome    &    0.032         &    0.036         &    0.030         \\
Observations    &   14,041         &    4,961         &    9,053         \\
\(R^2\)         &    0.099         &    0.146         &    0.114         \\
\bottomrule
\end{tabular}
}
\vspace{0.4em}
\begin{minipage}{16cm}
\footnotesize\singlespacing
\textit{Note}: Robust standard errors in parentheses are clustered at the product-category level, and the p-values are $^{*}p<0.10$, $^{**}p<0.05$, and $^{***}p<0.01$. Columns 1-3 report estimates from venture-level regressions in which the outcome is log venture capital financing raised within one year after first launch. The sample includes ventures surviving one year after first launch. All specifications include product category and year-quarter fixed effects, as well as venture controls, including fixed effects for founding team size, website creation month, and founder age group, gender, education, and employment background, and the log number of same-quarter new ventures launched in the same product category.
\end{minipage}
\end{table}

\end{document}


\maketitle
\singlespacing

\begin{center}
\end{center}

\appendix
\renewcommand{\thefigure}{A\arabic{figure}}
\renewcommand{\thetable}{A\arabic{table}}
\setcounter{figure}{0}
\setcounter{table}{0}

\section{Appendix Figures and Tables}

\begin{figure}[H]
    \centering
    \caption{Visualizations of Cluster-Based Structure of Product Categories}
    \label{fig:clustersadd}

    \begin{minipage}[t]{0.48\textwidth}
        \centering
        Panel (A): Product Categories as Hierarchical Clusters\\[2pt]
        \includegraphics[
            width=0.9\textwidth,
            trim=1cm 0.5cm 1cm 1.5cm,
            clip
        ]{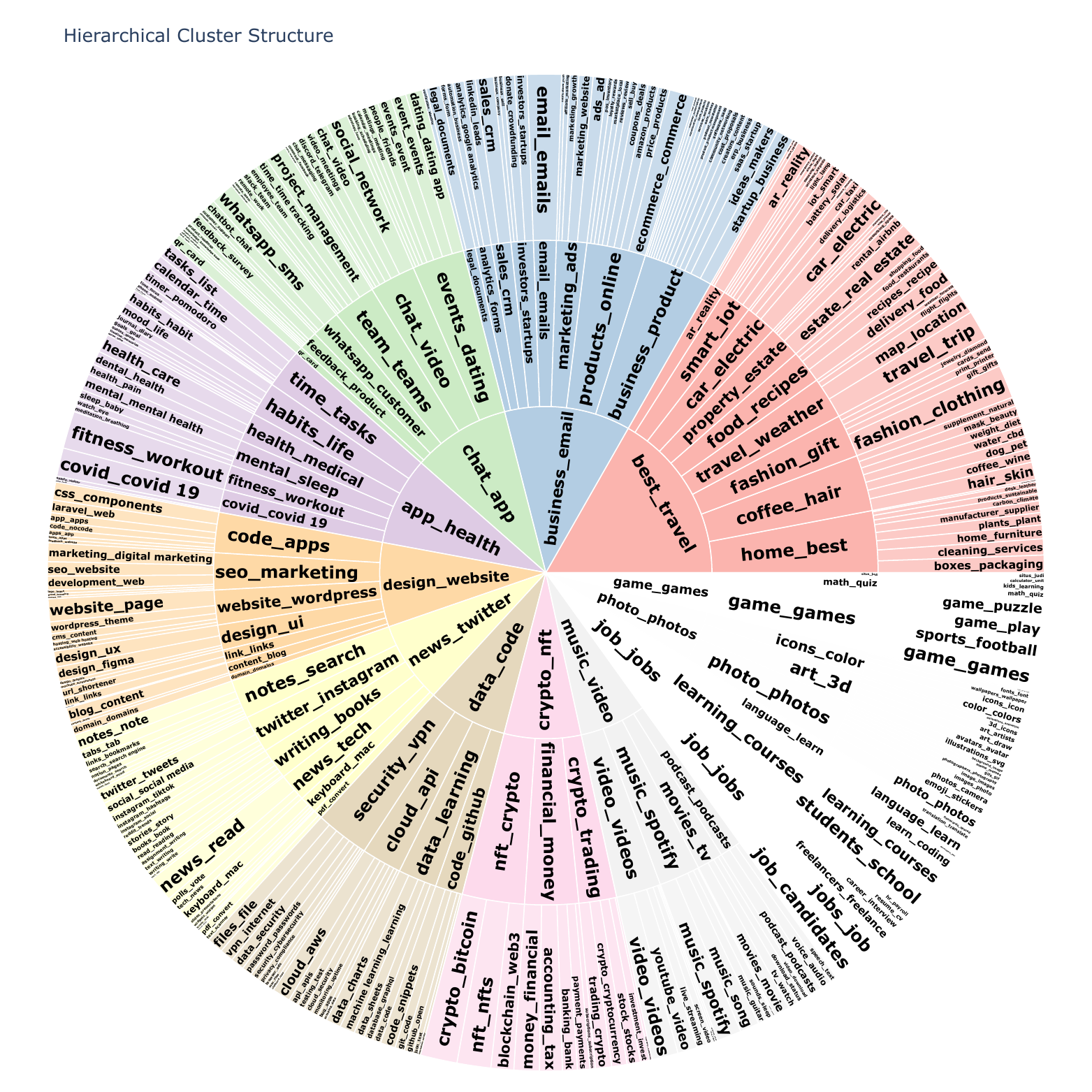}
    \end{minipage}
    \hfill
    \begin{minipage}[t]{0.48\textwidth}
        \centering
        Panel (B): 2D Projection of the Product Space\\[2pt]
        \includegraphics[
            width=0.9\textwidth,
            trim=1cm 1cm 1cm 2cm,
            clip
        ]{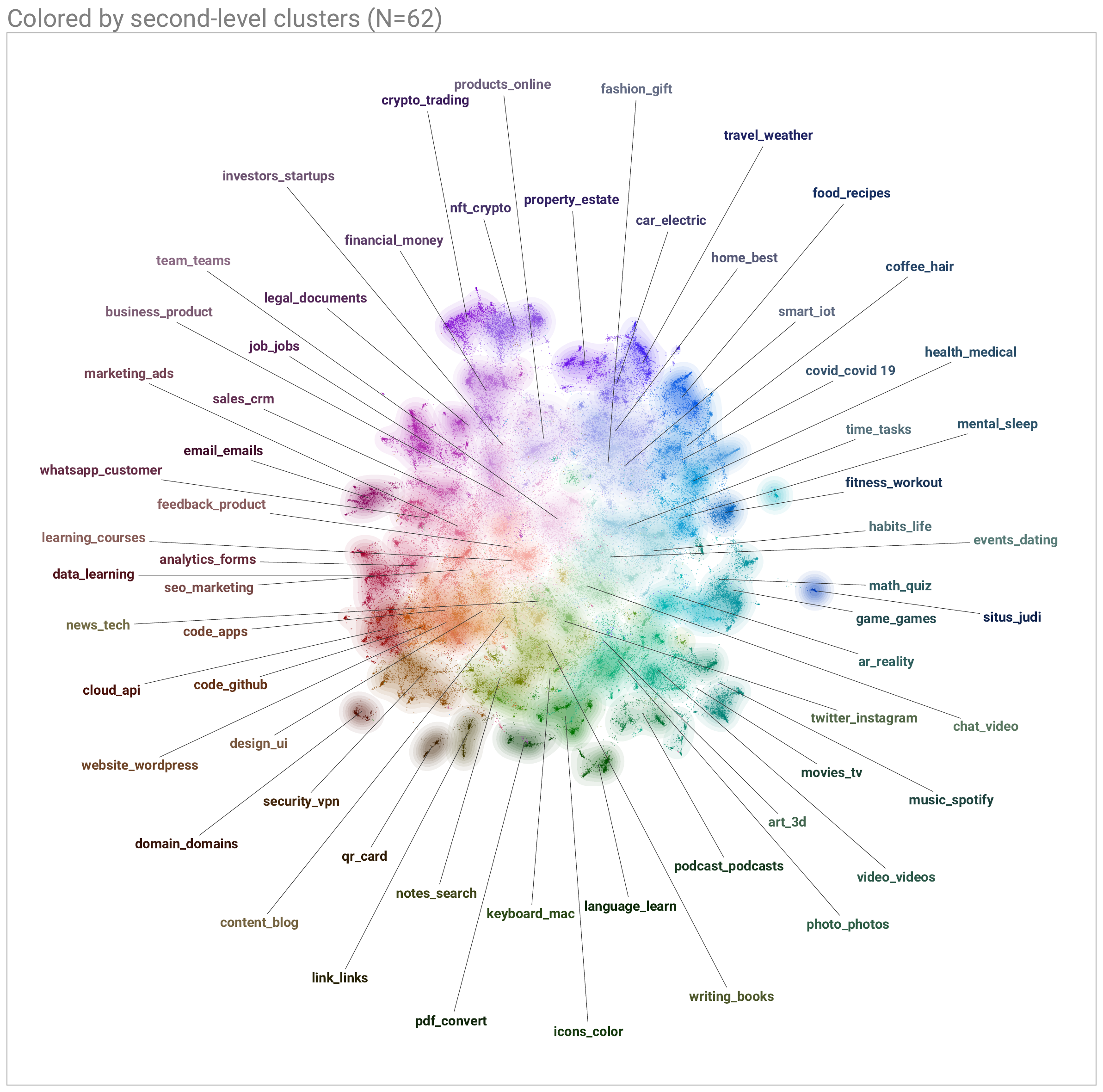}
    \end{minipage}

    \vspace{1em}

    \begin{minipage}{0.95\textwidth}
        \singlespacing
        \footnotesize
        \textit{Note}: Panel (A) visualizes the product-category hierarchy as a sunburst chart at three levels: Level 1 ($N=12$), Level 2 ($N=62$), and Level 3 ($N=310$). These categories are constructed from the linkage matrix of the hierarchical clustering algorithm trained on textual descriptions of products launched between 2019Q1 and 2022Q2. Panel (B) visualizes a two-dimensional projection of the Level-2 clusters ($N=62$). Each label is placed at the tip of an arrow extending outward from the corresponding cluster center. The final product categories are based on Level 3 ($N=310$) clusters.
    \end{minipage}
\end{figure}

\begin{figure}[H]
\caption{Descriptive Patterns in Product Categories}\label{appfig:categories}
\centering
\begin{minipage}{0.9\textwidth}
\centering
Panel (A): Product Category Size Distribution\\[2pt]
\includegraphics[width=0.8\textwidth]{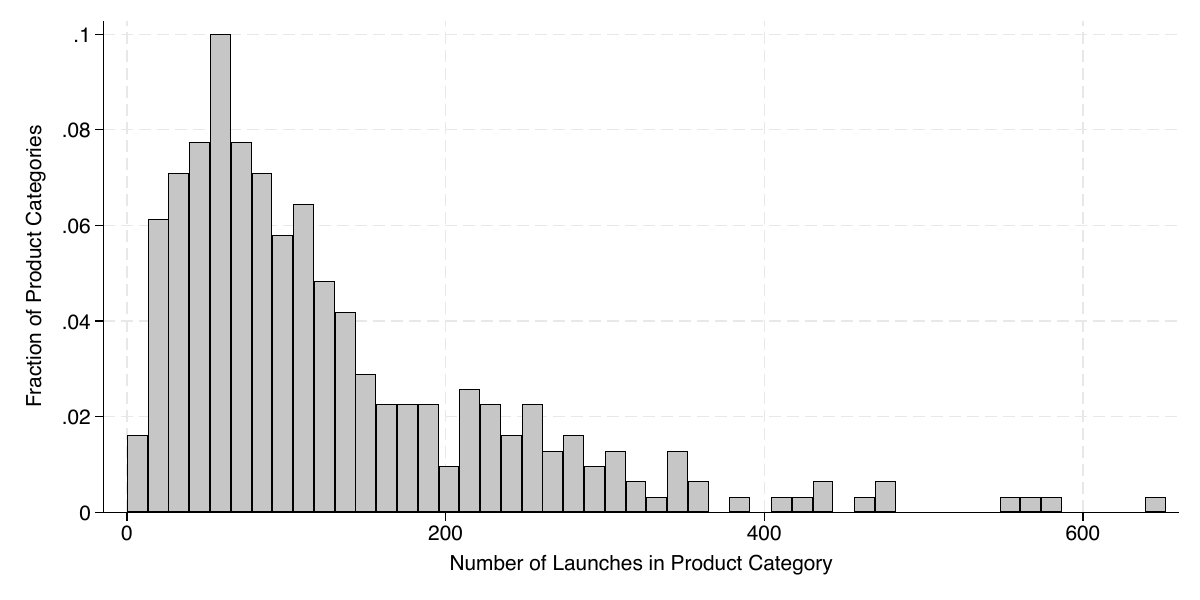}
\end{minipage}
\hfill\vspace{1em}
\begin{minipage}{0.9\textwidth}
\centering
Panel (B): Quarterly Number of New Venture Launches by Treatment Status\\[2pt]
\includegraphics[width=0.8\textwidth]{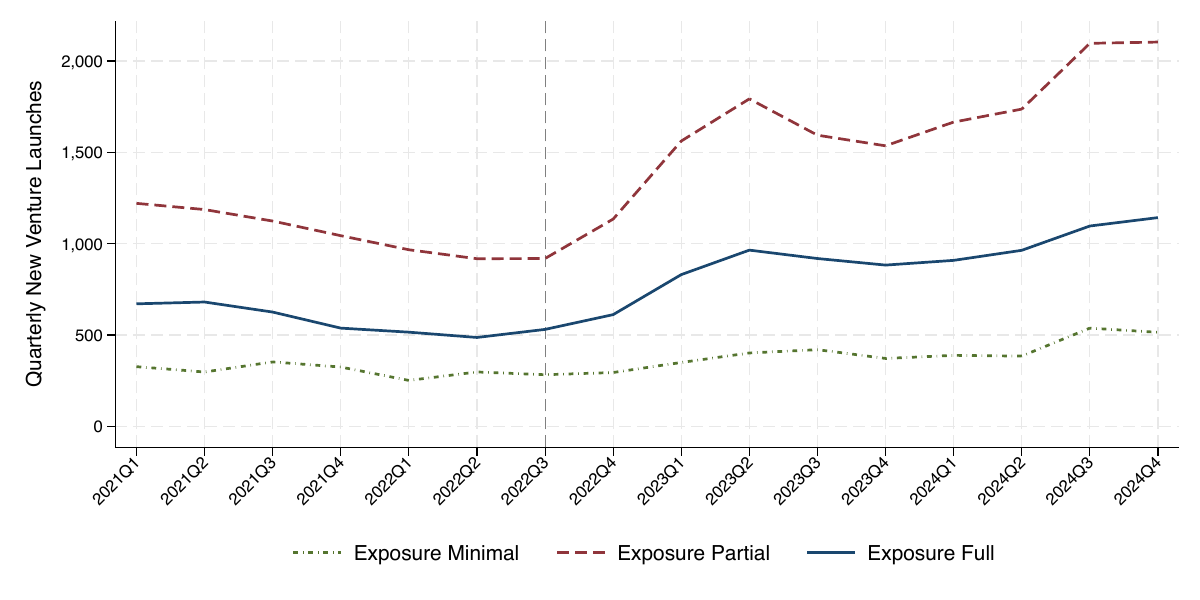}
\end{minipage}
\vspace{0.5em}
\begin{minipage}{0.9\textwidth}
\singlespacing\footnotesize \textit{Note}: Panel (A) shows the distribution of ventures in the sample across product categories. Panel (B) shows the quarterly total number of new venture launches across product categories with full, partial, and minimal exposure to AI coding, respectively.
\end{minipage}
\end{figure}

\begin{table}[H]
\caption{System Prompt for Classifying Products' Exposure to AI Coding}
\label{apptab:vibecodingprompt}
\end{table}
\noindent\textbf{Full prompt generated by Claude Opus 4.6:}
\smallskip
\begin{mdframed}[style=codebox]
\footnotesize\ttfamily\raggedright
SYSTEM\_PROMPT = \char`\"{}\char`\"{}\char`\"{}You are an expert software architect assessing whether products can be built using "vibe coding"---an AI-assisted development approach where a developer describes what they want in natural language and iterates with an AI coding assistant (like Claude, Cursor, or GitHub Copilot) without deep expertise in the underlying technologies.\par
\medskip
Your task: Given a product description, classify how much of the product can realistically be vibe-coded by a non-expert developer working with AI assistance.\par
\medskip
\#\# Classification Scale\par
\medskip
Rate the product on this 5-point scale:\par
\medskip
\#\#\# 4 - FULLY VIBE-CODABLE\par
The entire product can be built through AI-assisted iteration by someone without deep technical expertise.\par
\medskip
Characteristics:\par
- Standard CRUD operations\par
- Common UI patterns (forms, lists, dashboards, landing pages)\par
- Well-documented third-party APIs with good SDKs\par
- Static or simple dynamic content\par
- Standard authentication (using existing providers like Auth0, Clerk, Firebase Auth)\par
- Basic database operations with established ORMs\par
- No real-time requirements beyond simple polling\par
- Deployment on managed platforms (Vercel, Netlify, Railway)\par
\medskip
Examples: Portfolio sites, blogs, simple e-commerce storefronts, waitlist pages, internal dashboards, survey tools, basic SaaS landing pages with Stripe checkout\par
\medskip
\#\#\# 3 - MOSTLY VIBE-CODABLE\par
80-90\% can be vibe-coded, but 1-2 components require careful implementation or expert review.\par
\medskip
Characteristics:\par
- Everything in Level 4, plus:\par
- Moderately complex state management\par
- Multiple third-party integrations that must work together\par
- Basic real-time features (chat, notifications via services like Pusher/Ably)\par
- File upload/processing with size limits\par
- Role-based access control\par
- Search functionality (using Algolia, ElasticSearch-as-a-service)\par
- Email/SMS workflows via providers\par
\medskip
Risk areas that need expert review: Data model design, API rate limiting, basic security review\par
\medskip
Examples: Project management tools, CRM systems, booking/scheduling platforms, multi-tenant SaaS applications, marketplaces with standard features\par
\medskip
\#\#\# 2 - PARTIALLY VIBE-CODABLE\par
50-70\% can be vibe-coded. Core functionality requires traditional development expertise, but significant scaffolding, UI, and auxiliary features can be AI-assisted.\par
\medskip
Characteristics:\par
- Complex business logic with edge cases\par
- Custom integrations without good SDKs\par
- Performance-sensitive operations\par
- Complex data pipelines or ETL\par
- Advanced caching strategies\par
- Custom real-time synchronization (beyond simple pub/sub)\par
- Payment systems beyond basic checkout (escrow, splits, subscriptions with proration)\par
- Compliance requirements (GDPR data handling, audit logs)\par
\medskip
What CAN be vibe-coded: Admin panels, user-facing UI, standard API endpoints, documentation, test scaffolding, DevOps configuration\par
\medskip
Examples: Analytics platforms, workflow automation tools, complex e-commerce with inventory management, fintech applications (non-core), content management systems with custom publishing workflows\par
\medskip
\#\#\# 1 - MINIMALLY VIBE-CODABLE\par
Only 20-40\% can be vibe-coded. Most functionality requires deep technical expertise.\par
\medskip
Characteristics:\par
- Real-time systems with strict latency requirements\par
- Custom ML model training/deployment\par
- Complex distributed systems\par
- Video/audio processing pipelines\par
- Custom cryptographic implementations\par
- High-throughput data processing\par
- Game engines or physics simulations\par
- IoT device communication protocols\par
\medskip
What CAN be vibe-coded: Landing pages, documentation sites, admin interfaces, simple monitoring dashboards, marketing materials\par
\medskip
Examples: Video conferencing platforms, multiplayer game backends, algorithmic trading systems, real-time collaboration tools (like Figma), custom CDNs, recommendation engines\par
\medskip
\#\#\# 0 - NOT VIBE-CODABLE\par
Less than 20\% of meaningful functionality can be vibe-coded. Requires specialized expertise throughout.\par
\medskip
Characteristics:\par
- Novel algorithms or research implementations\par
- Hardware/firmware integration\par
- Safety-critical systems\par
- Custom database engines or protocols\par
- Compiler/language development\par
- Operating system components\par
- Medical device software\par
- Aerospace/automotive control systems\par
- High-frequency trading core\par
- Custom encryption/security protocols\par
\medskip
Examples: Self-driving car software, medical imaging analysis, chip design tools, quantum computing frameworks, blockchain consensus mechanisms, real-time operating systems\par
\medskip
\#\# Assessment Framework\par
\medskip
When evaluating, consider these dimensions:\par
1. Technical Complexity\par
2. Pattern Availability\par
3. Integration Complexity\par
4. Performance Requirements\par
5. Safety/Security Criticality\par
6. Regulatory Requirements\par
7. State Management Complexity\par
8. Real-time Requirements\par
\medskip
\#\# Output Format\par
\medskip
Respond with ONLY this JSON structure:\par
\medskip
\{"classification": <0|1|2|3|4>, "label": "<NOT\_VIBE\_CODABLE|MINIMALLY|PARTIALLY|MOSTLY|FULLY>", "confidence": "<HIGH|MEDIUM|LOW>", "reasoning": \{"vibe\_codable\_components": ["<list>"], "expert\_required\_components": ["<list>"], "key\_risk\_factors": ["<list>"], "key\_enablers": ["<list>"]\}, "summary": "<1-2 sentence explanation>"\}\par
\medskip
\#\# Important Guidelines\par
\medskip
1. Assume modern AI tools (2024-2025 capabilities).\par
2. Assume managed services are available.\par
3. Consider the 80/20 rule.\par
4. Don't penalize for scope.\par
5. Do penalize for depth.\par
6. Assume the developer is intelligent but not specialized.\par
7. Consider maintenance.\char`\"{}\char`\"{}\char`\"{}
\end{mdframed}

\begin{table}[H]
  \caption{Mapping the 0–4 Vibe-codability Scale to Exposure Intensity}
  \label{apptab:vibecodability}
  \centering
  \begin{tabular}{|l|c|l|}
  \hline
  \textbf{Intensity} & \textbf{Scale} & \textbf{Explanation} \\
  \hline
  2 (Full) & 4 & Standard patterns, AI can handle end-to-end \\
  \hline
  \multirow{2}{*}{1 (Partial)} & 3 & 1--2 components need traditional dev \\
   & 2 & Core logic needs expertise, but scaffolding/UI can be vibe-coded \\
  \hline
  \multirow{2}{*}{0 (Minimal)} & 1 & Only boilerplate can be AI-generated \\
   & 0 & Novel algorithms, hardware, real-time, security-critical \\
  \hline
  \end{tabular}
      \begin{minipage}{16cm}
      \vspace{-0.3cm}
      \singlespacing\footnotesize \textit{Note}: Product segments with a raw score of 4 are classified as fully exposed (\texttt{ExposureIntensity}=2), those with raw scores of 2 or 3 as partially exposed (\texttt{ExposureIntensity}=1), and those with raw scores of 0 or 1 as the minimally exposed (\texttt{ExposureIntensity}=0).
      \end{minipage}
  \end{table}

\begin{table}[H]
\centering
\caption{Timeline of Major GenAI Coding Tool and Platform Launches}
\small
    \begin{tabular}{|l|l|c|}
\hline
\textbf{Company} & \textbf{Feature} & \textbf{Launch Date} \\
\hline
Replit     & \textit{``Generate Code''} & Jun 2022 \\
\hline
GitHub     & \textit{GitHub Copilot (Public)} & Jun 2022 \\
\hline
Replit     & \textit{Replit Ghostwriter} & Oct 2022 \\
\hline
Windsurf   & \textit{Codeium Beta Launch} & Oct 2022 \\
\hline
Cursor     & \textit{Cursor Initial public release: AI-first IDE} & Mar 2023 \\
\hline
Framer AI  & \textit{``Start with AI'' generative page feature} & Jun 2023 \\
\hline
Softr      & \textit{AI app generator} & Oct 2023 \\
\hline
Bubble AI  & \textit{AI app generator} & Oct 2023 \\
\hline
Vercel     & \textit{Vercel v0: Generative UI} & Oct 2023 \\
\hline
Typedream  & \textit{Typedream AI} & Feb 2024 \\
\hline
Replit     & \textit{Replit Agent} & Sep 2024 \\
\hline
Bolt       & \textit{Bolt.new: AI prompt-to-app feature} & Oct 2024 \\
\hline
Lovable    & \textit{Lovable public launch} & Nov 2024 \\
\hline
Windsurf   & \textit{Windsurf Editor: the first agentic IDE} & Nov 2024 \\
\hline
\end{tabular}

    \begin{minipage}{14cm}
    \vspace{-0.3cm}
    \singlespacing\footnotesize \textit{Note}: This table shows the timeline of major launches by GenAI coding tools and platforms.
    \end{minipage}
\label{apptab:tools}
\end{table}

\begin{landscape}
\begin{table}[htbp]\centering
\caption{Effect of AI Coding Exposure on Share of New Ventures with Founder Characteristics}\label{apptab:selection}
\def\sym#1{\makebox[0pt][l]{\textsuperscript{#1}}}
\begin{tabular}{l*{7}{>{\centering\arraybackslash}p{2cm}}}
\toprule
                &\multicolumn{1}{c}{Solo}&\multicolumn{1}{c}{Serial}&\multicolumn{1}{c}{Hub Empl.}&\multicolumn{1}{c}{Top Eng. Edu.}&\multicolumn{1}{c}{Female}&\multicolumn{1}{c}{Age}&\multicolumn{1}{c}{Msc Deg.}\\\cmidrule(lr){2-2}\cmidrule(lr){3-3}\cmidrule(lr){4-4}\cmidrule(lr){5-5}\cmidrule(lr){6-6}\cmidrule(lr){7-7}\cmidrule(lr){8-8}
                &\multicolumn{1}{c}{(1)}         &\multicolumn{1}{c}{(2)}         &\multicolumn{1}{c}{(3)}         &\multicolumn{1}{c}{(4)}         &\multicolumn{1}{c}{(5)}         &\multicolumn{1}{c}{(6)}         &\multicolumn{1}{c}{(7)}         \\
\midrule
Exposure Any $\times$ Post&   -0.005         &    0.002         &    0.041\sym{*}  &    0.030\sym{**} &   -0.036\sym{*}  &   -0.208         &    0.009         \\
                &  (0.017)         &  (0.023)         &  (0.025)         &  (0.012)         &  (0.019)         &  (0.345)         &  (0.023)         \\
\addlinespace
Log (\# New Ventures)&    0.007         &   -0.005         &    0.003         &    0.012\sym{***}&   -0.002         &   -0.152         &   -0.008         \\
                &  (0.009)         &  (0.011)         &  (0.011)         &  (0.004)         &  (0.008)         &  (0.155)         &  (0.011)         \\
Product Category FEs&      Yes         &      Yes         &      Yes         &      Yes         &      Yes         &      Yes         &      Yes         \\
Launch YQ FEs   &      Yes         &      Yes         &      Yes         &      Yes         &      Yes         &      Yes         &      Yes         \\
\midrule
Avg. Outcome    &    0.806         &    0.352         &    0.318         &    0.055         &    0.168         &   27.784         &    0.365         \\
Observations    &    4,682         &    4,682         &    4,682         &    4,682         &    4,682         &    4,682         &    4,682         \\
\(R^2\)         &    0.176         &    0.134         &    0.112         &    0.107         &    0.114         &    0.101         &    0.112         \\
\bottomrule
\end{tabular}
\vspace{0.4em}
\begin{minipage}{22cm}
\footnotesize\singlespacing \textit{Note}: Robust standard errors in parentheses are clustered at the product-category level, and the p-values are  $^{*}p<0.10$, $^{**}p<0.05$, $^{***}p<0.01$. Each column reports estimates from product-category-by-quarter regressions with an outcome variable equal to the mean of a founder-related characteristic across launching ventures: indicator of solo founder in column (1); indicator of at least one serial founder in column (2); indicator of at least one founder recently employed in a leading startup ecosystem -- Density Leaders in Dealroom's Global Tech Ecosystem Index 2026 -- in column (3); indicator of at least one founder with bachelor's or higher degree from a top engineering university -- Top 20 in Engineering and Technology field in the 2026 QS World University Rankings -- in column (4); indicator of at least one female founder in column (5); median founder's age in column (6); and indicator of at least one founder with a master's or higher degree in column (7). All specifications include product category and year-quarter fixed effects.
\end{minipage}
\end{table}

\end{landscape}

\begin{table}[htbp]\centering
\def\sym#1{\ifmmode^{#1}\else\(^{#1}\)\fi}
\caption{Moderating Effects on Sample Excluding Founders with Top Engineering Education}\label{tab:notqs20}
\small
{
\begin{tabular}{l*{6}{c}}
\toprule
                &\multicolumn{3}{c}{Survival 1 Year Post}                &\multicolumn{3}{c}{Log (New Funding \textdollar{} Mil)} \\\cmidrule(lr){2-4}\cmidrule(lr){5-7}
Founder Experience:         &\multicolumn{1}{c}{All}&\multicolumn{1}{c}{Sftw=1}&\multicolumn{1}{c}{Sftw=0}&\multicolumn{1}{c}{All}&\multicolumn{1}{c}{Sftw=1}&\multicolumn{1}{c}{Sftw=0}\\
                &\multicolumn{1}{c}{(1)}&\multicolumn{1}{c}{(2)}&\multicolumn{1}{c}{(3)}&\multicolumn{1}{c}{(4)}&\multicolumn{1}{c}{(5)}&\multicolumn{1}{c}{(6)}\\
\midrule
Expo. Partial $\times$ Post&   -0.060\sym{***}&    0.024         &   -0.059\sym{***}&    0.012         &    0.106\sym{***}&    0.005         \\
                &  (0.020)         &  (0.034)         &  (0.021)         &  (0.018)         &  (0.034)         &  (0.019)         \\
\addlinespace
Expo. Full $\times$ Post&   -0.065\sym{***}&   -0.026         &   -0.067\sym{***}&    0.045\sym{***}&    0.137\sym{***}&    0.038\sym{**} \\
                &  (0.022)         &  (0.035)         &  (0.021)         &  (0.017)         &  (0.034)         &  (0.017)         \\
\addlinespace
Sftw. Expr.     &    0.029         &                  &                  &    0.072\sym{**} &                  &                  \\
                &  (0.026)         &                  &                  &  (0.029)         &                  &                  \\
\addlinespace
Post $\times$ Sftw. Expr.&   -0.069\sym{**} &                  &                  &   -0.079\sym{**} &                  &                  \\
                &  (0.033)         &                  &                  &  (0.032)         &                  &                  \\
\addlinespace
Expo. Partial $\times$ Sftw. Expr.&   -0.074\sym{**} &                  &                  &   -0.078\sym{**} &                  &                  \\
                &  (0.030)         &                  &                  &  (0.034)         &                  &                  \\
\addlinespace
Expo. Full $\times$ Sftw. Expr.&   -0.016         &                  &                  &   -0.065\sym{**} &                  &                  \\
                &  (0.031)         &                  &                  &  (0.032)         &                  &                  \\
\addlinespace
Expo. Partial $\times$ Post $\times$ Sftw. Expr.&    0.088\sym{**} &                  &                  &    0.080\sym{**} &                  &                  \\
                &  (0.037)         &                  &                  &  (0.036)         &                  &                  \\
\addlinespace
Expo. Full $\times$ Post $\times$ Sftw. Expr.&    0.040         &                  &                  &    0.070\sym{**} &                  &                  \\
                &  (0.038)         &                  &                  &  (0.034)         &                  &                  \\
Venture Controls&      Yes         &      Yes         &      Yes         &      Yes         &      Yes         &      Yes         \\
Product Category FEs&      Yes         &      Yes         &      Yes         &      Yes         &      Yes         &      Yes         \\
Launch YQ FEs   &      Yes         &      Yes         &      Yes         &      Yes         &      Yes         &      Yes         \\
\midrule
Avg. Outcome    &    0.420         &    0.418         &    0.421         &    0.027         &    0.031         &    0.025         \\
Observations    &   31,275         &   11,026         &   20,240         &   13,137         &    4,592         &    8,518         \\
\(R^2\)         &    0.084         &    0.119         &    0.087         &    0.095         &    0.161         &    0.109         \\
\bottomrule
\end{tabular}
}
\vspace{0.4em}
\begin{minipage}{17cm}
\footnotesize\singlespacing
\textit{Note}: Robust standard errors in parentheses are clustered at the product-category level, and the p-values are $^{*}p<0.10$, $^{**}p<0.05$, and $^{***}p<0.01$. Columns 1-3 report estimates from venture-level regressions in which the outcome is an indicator for whether the venture survived one year after first launch. Columns 4-6 report estimates from venture-level regressions in which the outcome is log venture capital financing raised within one year after first launch. The sample includes all ventures in the final sample except those with at least one founder holding a bachelor's degree or higher from a top engineering university, defined as a Top 20 university in Engineering and Technology in the 2026 QS World University Rankings. All specifications include product category and year-quarter fixed effects, as well as venture controls, including fixed effects for founding team size, website creation month, and founder age group, gender, education, and employment background, and the log number of same-quarter new ventures launched in the same product category.
\end{minipage}
\end{table}

\begin{table}[htbp]\centering
\def\sym#1{\ifmmode^{#1}\else\(^{#1}\)\fi}
\caption{Robustness Test with Alternative Moderators}\label{tab:alt}
\footnotesize
{
\begin{tabular}{l*{4}{c}}
\toprule
                &\multicolumn{2}{c}{Survival 1 Year Post}&\multicolumn{2}{c}{Log (New Funding \textdollar{} Mil)}\\\cmidrule(lr){2-3}\cmidrule(lr){4-5}
Sample:         &\multicolumn{1}{c}{All}&\multicolumn{1}{c}{All}&\multicolumn{1}{c}{Survivor}&\multicolumn{1}{c}{Survivor}\\
                &\multicolumn{1}{c}{(1)}&\multicolumn{1}{c}{(2)}&\multicolumn{1}{c}{(3)}&\multicolumn{1}{c}{(4)}\\
\midrule
Expo. Partial $\times$ Post&   -0.028         &   -0.027         &    0.040\sym{**} &    0.028\sym{*}  \\
                &  (0.019)         &  (0.023)         &  (0.019)         &  (0.017)         \\
\addlinespace
Expo. Full $\times$ Post&   -0.049\sym{**} &   -0.050\sym{**} &    0.071\sym{***}&    0.041\sym{**} \\
                &  (0.019)         &  (0.024)         &  (0.017)         &  (0.017)         \\
\addlinespace
Top Eng. Edu.   &    0.004         &                  &    0.039         &                  \\
                &  (0.045)         &                  &  (0.101)         &                  \\
\addlinespace
Post $\times$ Top Eng. Edu.&    0.015         &                  &    0.021         &                  \\
                &  (0.052)         &                  &  (0.118)         &                  \\
\addlinespace
Expo. Partial $\times$ Top Eng. Edu.&    0.029         &                  &    0.065         &                  \\
                &  (0.052)         &                  &  (0.110)         &                  \\
\addlinespace
Expo. Full $\times$ Top Eng. Edu.&    0.025         &                  &    0.079         &                  \\
                &  (0.061)         &                  &  (0.113)         &                  \\
\addlinespace
Expo. Partial $\times$ Post $\times$ Top Eng. Edu.&   -0.035         &                  &   -0.085         &                  \\
                &  (0.060)         &                  &  (0.129)         &                  \\
\addlinespace
Expo. Full $\times$ Post $\times$ Top Eng. Edu.&   -0.011         &                  &   -0.123         &                  \\
                &  (0.072)         &                  &  (0.131)         &                  \\
\addlinespace
Hub Empl        &                  &    0.017         &                  &    0.093\sym{*}  \\
                &                  &  (0.026)         &                  &  (0.048)         \\
\addlinespace
Post $\times$ Hub Empl&                  &   -0.005         &                  &   -0.070         \\
                &                  &  (0.038)         &                  &  (0.055)         \\
\addlinespace
Expo. Partial $\times$ Hub Empl&                  &    0.017         &                  &   -0.014         \\
                &                  &  (0.029)         &                  &  (0.053)         \\
\addlinespace
Expo. Full $\times$ Hub Empl&                  &    0.023         &                  &   -0.074         \\
                &                  &  (0.032)         &                  &  (0.051)         \\
\addlinespace
Expo. Partial $\times$ Post $\times$ Hub Empl&                  &   -0.010         &                  &    0.013         \\
                &                  &  (0.041)         &                  &  (0.059)         \\
\addlinespace
Expo. Full $\times$ Post $\times$ Hub Empl&                  &    0.000         &                  &    0.054         \\
                &                  &  (0.043)         &                  &  (0.057)         \\
Venture Controls&      Yes         &      Yes         &      Yes         &      Yes         \\
Product Category FEs&      Yes         &      Yes         &      Yes         &      Yes         \\
Launch YQ FEs   &      Yes         &      Yes         &      Yes         &      Yes         \\
\midrule
Avg. Outcome    &    0.424         &    0.424         &    0.032         &    0.032         \\
Observations    &   33,155         &   33,155         &   14,041         &   14,041         \\
\(R^2\)         &    0.083         &    0.084         &    0.097         &    0.099         \\
\bottomrule
\end{tabular}
}
\vspace{0.4em}
\begin{minipage}{15cm}
\footnotesize\singlespacing
\textit{Note}: Robust standard errors in parentheses are clustered at the product-category level, and the p-values are $^{*}p<0.10$, $^{**}p<0.05$, and $^{***}p<0.01$. All columns report estimates from venture-level regressions, and control for product category and year-quarter fixed effects, as well as venture controls, including fixed effects for founding team size, website creation month, and founder age group, gender, education, and employment background, and the log number of same-quarter new ventures launched in the same product category.
\end{minipage}
\end{table}

\clearpage
\begin{table}[H]
\centering
\caption{Products' AI-Native Likelihood Scoring Prompt and Measure Distribution}
\label{tab:ainativeprompt}

\begin{subfigure}{\linewidth}
\caption{AI-Native Likelihood Scoring Prompt}
\centering
\begin{mdframed}[
  linewidth=0.5pt,
  linecolor=black,
  backgroundcolor=gray!5,
  innertopmargin=8pt,
  innerbottommargin=8pt,
  innerleftmargin=10pt,
  innerrightmargin=10pt,
  roundcorner=4pt
]
\begin{lstlisting}[
  basicstyle=\ttfamily\small,
  breaklines=true,
  columns=fullflexible,
  keepspaces=true
]
You are scoring venture/product descriptions for how AI-native the product is.
Score the likelihood (0.0 to 1.0) that AI is the CORE of the product's value proposition, as opposed to a peripheral feature bolted onto a fundamentally non-AI product.
- Close to 1.0: the product's entire function is built around AI (e.g. an AI video generator, an LLM-powered coding assistant, an AI chatbot platform, an AI writing tool).
- Close to 0.0: AI is absent, or is merely a minor feature/enhancement on a non-AI product (e.g. a CRM with an added "AI summary" button, a delivery app whose core function is logistics but whose marketing mentions "AI-powered" recommendations).
- In between: genuine ambiguity about how central AI is to the product's core offering.
For each product below, output a likelihood score in the "scores" array of your response, one entry per product_id, using the exact product_id given.
Products:
<... followed by, for each product in the batch:>
product_id=<product_id>
description: <combined column text, verbatim, NaN filled as empty string>
\end{lstlisting}
\end{mdframed}
\end{subfigure}

\vspace{2em}

\begin{subfigure}{\linewidth}
\caption{Distribution of AI-Native Likelihood Scores}
\centering
\includegraphics[width=0.8\linewidth]{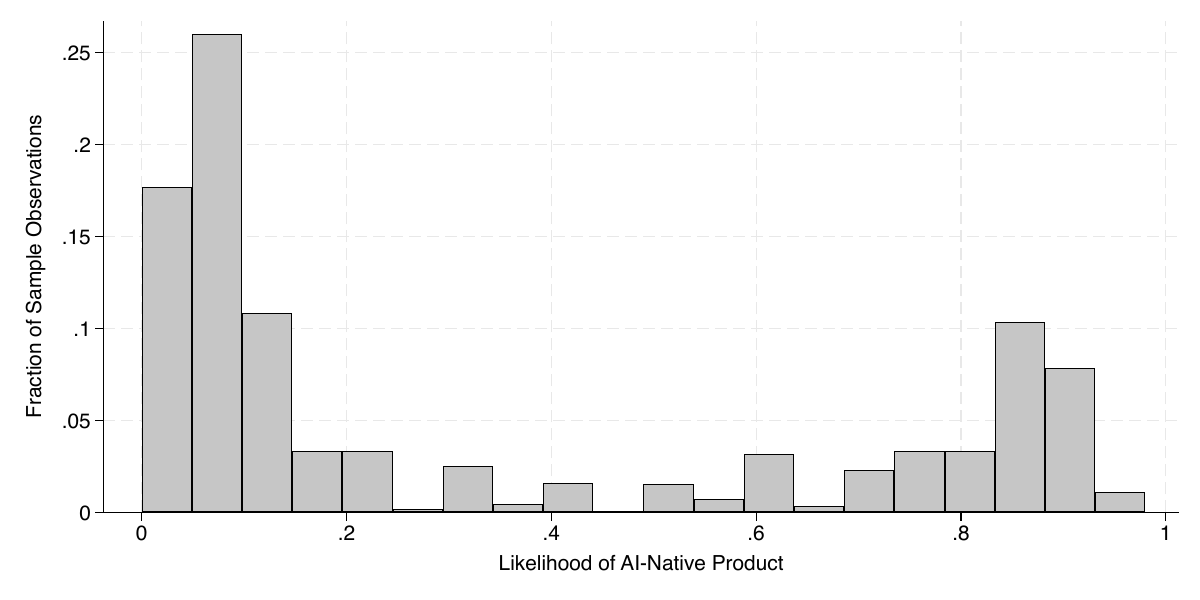}
\end{subfigure}

\end{table}

\begin{table}[htbp]\centering
\def\sym#1{\ifmmode^{#1}\else\(^{#1}\)\fi}
\caption{Moderating Effects}\label{tab:genai}
\small
{
\begin{tabular}{l*{3}{c}}
\toprule
                &\multicolumn{3}{c}{Likelihood of AI-Native Product}     \\\cmidrule(lr){2-4}
Founder Experience:         &\multicolumn{1}{c}{All}&\multicolumn{1}{c}{Sftw=1}&\multicolumn{1}{c}{Sftw=0}\\
                &\multicolumn{1}{c}{(1)}&\multicolumn{1}{c}{(2)}&\multicolumn{1}{c}{(3)}\\
\midrule
Expo. Partial $\times$ Post&    0.031         &    0.090\sym{***}&    0.030         \\
                &  (0.035)         &  (0.024)         &  (0.036)         \\
\addlinespace
Expo. Full $\times$ Post&    0.011         &    0.103\sym{***}&    0.009         \\
                &  (0.038)         &  (0.027)         &  (0.039)         \\
\addlinespace
Sftw. Expr.     &    0.036         &                  &                  \\
                &  (0.029)         &                  &                  \\
\addlinespace
Post $\times$ Sftw. Expr.&   -0.038         &                  &                  \\
                &  (0.031)         &                  &                  \\
\addlinespace
Expo. Partial $\times$ Sftw. Expr.&   -0.057\sym{*}  &                  &                  \\
                &  (0.030)         &                  &                  \\
\addlinespace
Expo. Full $\times$ Sftw. Expr.&   -0.063\sym{**} &                  &                  \\
                &  (0.031)         &                  &                  \\
\addlinespace
Expo. Partial $\times$ Post $\times$ Sftw. Expr.&    0.048         &                  &                  \\
                &  (0.033)         &                  &                  \\
\addlinespace
Expo. Full $\times$ Post $\times$ Sftw. Expr.&    0.091\sym{**} &                  &                  \\
                &  (0.037)         &                  &                  \\
Venture Controls&      Yes         &      Yes         &      Yes         \\
Product Segment FEs&      Yes         &      Yes         &      Yes         \\
Launch YQ FEs   &      Yes         &      Yes         &      Yes         \\
\midrule
Avg. Outcome    &    0.325         &    0.343         &    0.315         \\
Observations    &   14,041         &    4,961         &    9,053         \\
\(R^2\)         &    0.477         &    0.517         &    0.480         \\
\bottomrule
\end{tabular}
}
\vspace{0.4em}
\begin{minipage}{12.5cm}
\footnotesize\singlespacing
\textit{Note}: Robust standard errors in parentheses are clustered at the product-category level, and the p-values are $^{*}p<0.10$, $^{**}p<0.05$, and $^{***}p<0.01$. Columns 1-3 report estimates from venture-level regressions in which the outcome is a continuous measure of AI-native likelihood -- i.e., how likely AI is the core of the venture's first launch product's value proposition -- ranging from 0 to 1. All specifications include product category and year-quarter fixed effects, as well as venture controls, including fixed effects for founding team size, website creation month, and founder age group, gender, education, and employment background, and the log number of same-quarter new ventures launched in the same product category.
\end{minipage}
\end{table}

\section{Data Construction}

\subsection{Description of Data Sources} \label{appsec:match}

\textbf{Product Hunt.} Product Hunt is a global platform where new digital ventures from around the world announce their product launches, in order to gauge market traction, acquire early adopters, and validate their ideas \citep{cao2024sampling}. From the platform’s website, we collect detailed data on all product launches -- including official websites and textual descriptions -- and the venture teams behind these launches -- including each team member’s name, short biography, and social platform links (specifically, LinkedIn and X). We identify each venture’s \textit{initial} launch by linking their official websites’ registered domain across launches of the same underlying venture, and retaining only the venture’s first launch.

\textbf{Revelio Labs.} Revelio Labs is a comprehensive workforce database that aggregates hundreds of millions of public employment records, covering over 1.1 billion individuals and more than 20 million companies through LinkedIn profile data. These data on individual career histories and educational backgrounds enable us to measure various aspects of venture team members’ demographic profiles -- such as location, age, and gender -- and professional expertise -- such as software experience and engineering training -- prior to launching their ventures. The record matching for each venture team member from individuals’ user profiles on Product Hunt to Revelio Labs is performed using a combination of approaches involving both direct matching on social platform links and automated search-engine queries.\footnote{For detailed description of the record matching process, see Appendix \ref{appsec:sample}.}

\textbf{Semrush.} Semrush is a marketing analytics platform that combines clickstream data from hundreds of millions of Internet users worldwide with machine-learning methods to generate engagement metrics. Semrush’s data collection covers nearly all website domains with any observed user traffic at all, and their monthly website traffic data has been used in prior research to measure real outcomes of digital ventures such as survival and performance \citealp{koning2022experimentation}. The matching between other venture-level data sources and Semrush is performed using the venture’s registered domain, which uniquely identifies any digital venture with a standalone website.

\textbf{PitchBook.} PitchBook is a leading commercial database on global private capital market activities, with particularly comprehensive coverage of financing activities by technology startups and private companies that have raised financing rounds from venture capital and private equity investors. Since registered domain uniquely identifies any digital venture with a standalone website, we use it to match Pitchbook data to sample firms to measure each firm’s financing status and cumulative amount of financing raised.

\subsection{Product Hunt Sample Construction}

To examine how GenAI affects the entry and performance of nascent digital ventures, we collect large-scale data from Product Hunt, a global platform where early-stage ventures launch technology products, gauge market traction, acquire early users, and validate ideas \citep{cao2024sampling}. With roughly four million monthly visits, Product Hunt provides a suitable empirical context for studying AI coding because it covers information about ventures at an early stage, when founders are preparing for market entry but have not yet accumulated substantial users, which makes GenAI-enabled software development especially valuable for implementing entrepreneurial ideas and developing initial working products.

We collect rich data from Product Hunt on digital ventures’ first public launches. These data include detailed information on the launches themselves as well as ex ante characteristics of ventures and their founders. Specifically, we observe venture-level information, such as website domains, product categories, and textual descriptions, and founder-level information based on listed product ``makers,’’ such as names, headlines, and social links. Because our sample consists of first-time launches observed close to founding, the maker team listed on the launch page provides a reasonable proxy for the founding team. These data allow us to augment our sample by linking founders to Revelio Labs and ventures to Semrush, CrunchBase, and PitchBook.

For each founder, we use the information listed on their Product Hunt account—including name, headline, and social links, especially LinkedIn and Twitter—to identify the best match to a Revelio Labs profile, either directly or through CrunchBase. The detailed matching procedure is described in Appendix Section \ref{appsec:sample}. Revelio Labs is a comprehensive workforce database that aggregates hundreds of millions of public employment records, covering over 1.1 billion individuals and more than 20 million companies through LinkedIn profile data. These individual-level job and education histories enable us to construct measures of founders’ demographics (e.g., gender, ethnicity, and age), career histories, and educational backgrounds.

For each venture, we use the \textit{registered domain} recorded at its first Product Hunt launch as the venture’s unique identifier. For ventures without a standalone domain that instead rely on a third-party host, such as a mobile app or Chrome extension (about 12\% of launches in our sample period), we use the full website path instead. Using venture’s registered domains, we collect monthly website visits from Semrush, a marketing analytics platform that combines clickstream data from hundreds of millions of Internet users worldwide with machine-learning methods to generate engagement metrics. Because Semrush covers nearly all domains with observed user activity, prior research commonly uses its traffic data to measure digital ventures’ customer acquisition and performance (e.g., \citealp{koning2022experimentation,cao2024sampling,conti2025selective}). Finally, to verify that our sample consists of entrepreneurial ventures, we match each launch to CrunchBase and PitchBook to obtain founding dates. Both databases provide structured information on companies, prominent individuals, and financing activity in technology entrepreneurship, with particularly strong coverage of venture-backed private firms.

To focus the analysis on entrepreneurial ventures undertaking first-time launches, we apply the following criteria. First, we exclude launch posts without listed makers, because venture-initiated prepared launches always identify their makers. Second, we exclude observations for which 50\% or more of the makers cannot be reliably matched to Revelio Labs. We remove podcast episodes, games, books, and other posts that are not entrepreneurial venture products. In addition, we exclude ventures hosted on third-party platforms without standalone domains, except for mobile apps and Chrome extensions, which are legitimate software products. Finally, to exclude large incumbent firms, we drop organizations that launched more than 30 times, or had more than five makers, or already raised external financing before the launch, or were founded more than five years before their first launch according to CrunchBase, PitchBook, or LinkedIn, or had registered domains created more than four years before their first launch. These restrictions retain 82\% of all first-time launches by ventures for which the majority of founders can be identified in Revelio Labs.

\subsection{Sampling Criteria and Final Data} \label{appsec:sample}

Our final dataset is constructed by matching all launch posts on Product Hunt with makers' LinkedIn career histories using the Revelio Labs database. Our data construction recovers the information hierarchy of Product Hunt by tracing launch posts (\textit{PostId}) to products (\textit{ProductId}) and then to the parent organization (\textit{websiteDomain}\footnote{The only exceptions are platform complementors, in particular mobile apps and Chrome extensions.}). This process leaves over 100,000 launch posts from 2021 to 2025, which are further filtered to construct the final sample of new ventures as follows.

\begin{enumerate}
    \item \textbf{Fixing the unit of analysis at the venture level.} Our primary unit of analysis is a venture rather than a product or launch post, as our analyses focus on entrepreneurial firms and their founders rather than product lines within established companies. We focus on ventures also because more information is available at the organizational level by matching each venture's website domain to external databases to obtain additional variables.

    \item \textbf{Removing non-launch posts with no makers tagged.} To focus on firms' intended public launches rather than hunter-initiated product discoveries, we keep only launch posts that tag at least one maker. Posts without tagged makers are not firm-initiated but externally hunted, typically by unaffiliated users discovering and posting a product after it has already been widely adopted.

    \item \textbf{Consolidating information on the first launch and the overall firm.} We combine data from launch posts, products, and external databases linked to each parent organization to construct two types of variables for each organization:
    \begin{itemize}
        \item \textit{Time of the venture's first launch}: We identify the earliest timestamped launch post across all products sharing the same organization identifier, primarily the domain name, and treat it as the startup's launch date. Although a firm may launch multiple times under the same product, such as feature updates or new versions, or across distinct product lines, we use the first post on Product Hunt to mark when the firm conducted its initial public launch to acquire early users. This approach captures both the initial public launch and the time-to-launch measure, based on the website creation date, which typically precedes the initial public launch by months or years.

        \item \textit{Venture and founder characteristics}: We combine data from all products and launch posts linked to the same parent organization to construct venture-level variables. First, we use textual information from each product's tagline and description to characterize the product space. We use the embedding vectors of these texts and apply agglomerative hierarchical clustering to group the launches into mutually exclusive product segments.\footnote{Because ventures may pivot later, we define their initial product segment using information from the first launch. In most cases, a venture has a single product and a single launch, though some ventures have multiple products or additional launches, such as updates or new feature releases.} Second, we use the registered website domain as a unique venture identifier, which allows us to match ventures to external databases, including LinkedIn profiles, PitchBook, CrunchBase, and Semrush. Third, we use tagged maker information, including names, social links, and headlines, to match makers to Revelio Labs. When the maker team is small enough, we can treat tagged makers as members of the founding team.
    \end{itemize}

    \item \textbf{Matching founders from Product Hunt makers to LinkedIn profiles.} We obtain additional founder information by matching makers' Product Hunt accounts to their LinkedIn profiles. Founders are identified from the makers tagged in each organization's first launch post. We then construct a founder--firm-level dataset and aggregate it to the firm level to create ex-ante founding-team measures, including the primary founder's demographics, location, age, and gender, work experience, and education history.
    \begin{itemize}
        \item \textit{Keeping only ventures with sufficient information on founders}: To ensure sufficiently accurate measurement of founding teams' educational and professional backgrounds, we retain only launches where more than 50\% of the founding team members, i.e., makers, can be reliably matched to Revelio Labs. This means that the match satisfies at least one of the following criteria: (1) the individual's work experience includes the launched venture, (2) the individual's Product Hunt profile lists their LinkedIn profile URL, (3) the individual can be linked to CrunchBase, either through the Twitter handle in their Product Hunt profile or through their full name and the launched venture's website domain, and the CrunchBase profile contains their LinkedIn profile URL, or (4) the individual is matched directly to a LinkedIn URL through their full name, i.e., the last name is not missing or initials only.
    \end{itemize}
\end{enumerate}

\subsection{Why Fine-Tuning the Embedding Model Is Not Necessary for Deriving Product Clusters}
\label{appsec:finetuning}

The product-clustering procedure relies on text embeddings derived from each product's textual description. Although recent open-weights embedding models are highly capable, a potential concern is that an off-the-shelf model may not fully capture domain-specific nuances in digital product descriptions. We therefore examined whether fine-tuning an embedding model on Product Hunt-specific data would improve the quality of the embeddings used to construct product clusters.

To construct training data for fine-tuning, we use Product Hunt's ``alternative products'' information. For each product, we collect its tagline, description, and, when available, the set of products listed as alternatives. Around July--August 2025, Product Hunt appears to have changed the way it identifies and displays alternative products. After this change, the platform reports up to ten alternative products for selected focal products, with rankings based on a combined relevance score. The score appears to combine three components: an embedding-based score, a category-based score, and a rating-based score, with the combined score calculated as
\[
    \text{combinedScore}
    =
    3 \times \text{embeddingScore}
    +
    2 \times \text{categoryScore}
    +
    1 \times \text{ratingScore}
\]
Because Product Hunt only displays the top ten alternatives for products with available alternatives, we cannot observe the full distribution of relevance scores or define a universal cutoff for positive matches. In addition, many products have no listed alternatives, and for some products the underlying reason for missing alternatives is unclear. We therefore treat the displayed alternatives as positive examples, while recognizing that they provide an incomplete but useful source of product-level similarity information.

Each product is represented by concatenating its tagline and description, ensuring that each non-missing component ends with a period. We retain only products with non-missing descriptions, because descriptions contain substantially richer and more accurate information than taglines alone. For each focal product with at least one listed alternative, we treat the listed alternatives as positive examples. We then mine hard negative examples from the full set of products, not only from products appearing in the alternatives data. Specifically, using embeddings from the base model, we identify products whose cosine similarity to the focal product falls within a relatively high similarity range, such as 0.80--0.90 or 0.80--0.85, but that are not listed as alternatives and do not share the same website domain as the focal product. For each anchor--positive pair, we randomly select up to five eligible hard negatives.

We standardize the resulting data into the triplet format commonly used in contrastive learning. Each training observation consists of an anchor product, a positive product, and a negative product. Thus, for each anchor--positive pair with up to five sampled negatives, we create multiple triplets:
\[
    (\text{anchor}, \text{positive}, \text{negative}_{1}), \quad
    (\text{anchor}, \text{positive}, \text{negative}_{2}), \quad \ldots
\]
This format encourages the model to place the anchor product closer to the positive alternative than to hard negative products that are textually similar but not identified by Product Hunt as alternatives.

We experimented with fine-tuning several large embedding models, including Qwen3-Embedding-4B, as well as large BGE and E5 models. Before fine-tuning, we examined whether the base embeddings already captured meaningful product similarity. Using the base Qwen3-Embedding-4B model, we compared the cosine similarity between each anchor product and its listed alternatives with the similarity between the anchor and other nearby products in the embedding space. The listed alternatives generally had higher cosine similarity than other nearby products, and many positive examples had similarities above 0.85. This pattern suggests that the off-the-shelf model already captures substantial Product Hunt-specific similarity structure. It also informed the hard-negative mining strategy, since negatives with similarities around or slightly below 0.85 are likely to be informative for contrastive learning.

We then fine-tuned Qwen3-Embedding-4B using the triplet data. To reduce memory requirements, training used BF16 precision. We used a learning rate of $5 \times 10^{-6}$ and a warmup ratio of 0.1. We held out 2\% of the triplet data as an evaluation sample, corresponding to approximately 5,000 observations, and used the remaining 98\% for training. Performance was evaluated using the same triplet-based cosine-similarity comparison metric used during fine-tuning. However, even after approximately 100 training iterations, evaluation performance was already worse than that of the base model without fine-tuning. This suggests that fine-tuning on the Product Hunt alternatives data was more likely to induce overfitting than to improve the embedding space used for clustering.

Given this result, we proceeded by comparing several off-the-shelf embedding models without fine-tuning. Specifically, we evaluated non-quantized Qwen3-Embedding-4B, Qwen3-Embedding-8B, and large BGE and E5 models. Because the Qwen models are large, we compute embeddings using FP16 precision rather than FP32 to reduce memory requirements; this choice is substantially more efficient and has minimal practical effect on embedding quality. The comparison uses the same triplet-evaluation data and cosine-similarity metric as in the fine-tuning evaluation. Qwen3-Embedding-4B slightly outperformed Qwen3-Embedding-8B, and both Qwen models substantially outperformed the large BGE and E5 models.

We therefore use Qwen3-Embedding-4B as the embedding model for constructing product representations. For each product, we generate an embedding vector from the concatenated tagline and description text, and use these embeddings as inputs to the clustering procedure. The evidence from the fine-tuning exercise suggests that additional task-specific fine-tuning is unnecessary for this application: the base Qwen3-Embedding-4B model already captures the relevant semantic structure of Product Hunt product descriptions, while fine-tuning on the available alternatives data reduces out-of-sample evaluation performance.

\singlespacing
\bibliographystyle{apalike}
\bibliography{references_appendix}